\pdfoutput=1 
\documentclass[
    aip,
    jcp,
    amsmath,amssymb,
    reprint
    ]{revtex4-1}
\usepackage{graphicx}
\usepackage{dcolumn}
\usepackage{bm}
\usepackage[utf8]{inputenc}
\usepackage[T1]{fontenc}
\usepackage{etoolbox}

\usepackage{color}
\usepackage{mathrsfs}
\usepackage{mathtools} 
\usepackage{algorithmic}
\usepackage{algorithm}

\makeatletter
\def\@email#1#2{%
 \endgroup
 \patchcmd{\titleblock@produce}
  {\frontmatter@RRAPformat}
  {\frontmatter@RRAPformat{\produce@RRAP{*#1\href{mailto:#2}{#2}}}\frontmatter@RRAPformat}
  {}{}
}%
\makeatother
\begin{document}

\preprint{AIP/123-QED}

\title{Low-Dimensional Projection of Reactive Islands in Chemical Reaction Dynamics
       Using a Supervised Dimensionality Reduction Method}

\author{Ryoichi Tanaka}
    \affiliation{Graduate School of Chemical Sciences and Engineering, Hokkaido University, Sapporo 060-8628, Japan}
\author{Yuta Mizuno}
    \affiliation{Graduate School of Chemical Sciences and Engineering, Hokkaido University, Sapporo 060-8628, Japan}
    \affiliation{Research Institute for Electronic Science, Hokkaido University, Sapporo 001-0020, Japan}
    \affiliation{Institute for Chemical Reaction Design and Discovery (WPI-ICReDD), Hokkaido University, Sapporo 001-0021, Japan}
    \email{mizuno@es.hokudai.ac.jp}
\author{Takuro Tsutsumi}
    \affiliation{Graduate School of Chemical Sciences and Engineering, Hokkaido University, Sapporo 060-8628, Japan}
    \affiliation{Department of Chemistry, Faculty of Science, Hokkaido University, Sapporo 060-0810, Japan}
\author{Mikito Toda}
    \affiliation{Graduate School of Information Science, University of Hyogo, Kobe, 650-0047, Japan}
    \affiliation{Research Institute for Electronic Science, Hokkaido University, Sapporo 001-0020, Japan}
    \affiliation{Faculty Division of Natural Sciences, Research Group of Physics, Nara Women’s University,  Nara 630‐8506, Japan}
\author{Tetsuya Taketsugu}
    \affiliation{Graduate School of Chemical Sciences and Engineering, Hokkaido University, Sapporo 060-8628, Japan}
    \affiliation{Institute for Chemical Reaction Design and Discovery (WPI-ICReDD), Hokkaido University, Sapporo 001-0021, Japan}
    \affiliation{Department of Chemistry, Faculty of Science, Hokkaido University, Sapporo 060-0810, Japan}
\author{Tamiki Komatsuzaki}
    \affiliation{Graduate School of Chemical Sciences and Engineering, Hokkaido University, Sapporo 060-8628, Japan}
    \affiliation{Research Institute for Electronic Science, Hokkaido University, Sapporo 001-0020, Japan}
    \affiliation{Institute for Chemical Reaction Design and Discovery (WPI-ICReDD), Hokkaido University, Sapporo 001-0021, Japan}
    \affiliation{SANKEN, Osaka University, Ibaraki 567-0047, Japan}

\date{\today}

\begin{abstract}
  Transition state theory is a standard framework for predicting the rate of a chemical reaction. Although the transition state theory has been successfully
  applied to numerous chemical reaction analyses, many experimental and theoretical studies
  have reported chemical reactions with a reactivity which cannot be explained by the transition state theory
  due to dynamic effects. Dynamical systems theory provides a theoretical framework for elucidating
  dynamical mechanisms of such chemical reactions. In particular, reactive islands are essential
  phase space structures revealing dynamical reaction patterns. However, the numerical computation
  of reactive islands in a reaction system of many degrees of freedom involves an intrinsic
  challenge---the curse of dimensionality. In this paper, we propose a dimensionality reduction
  algorithm for computing reactive islands in a reaction system of many degrees of freedom.
  Using the supervised principal component analysis, the proposed algorithm projects
  reactive islands into a low-dimensional phase space with preserving the dynamical information
  on reactivity as much as possible. The effectiveness of the proposed algorithm is examined
  by numerical experiments for Hénon-Heiles systems extended to many degrees of freedom.
  The numerical results indicate that our proposed algorithm is effective in terms of
  the quality of reactivity prediction and the clearness of the boundaries of projected
  reactive islands. The proposed algorithm is a promising elemental technology for
  practical applications of dynamical systems analysis to real chemical systems.

\end{abstract}

\maketitle

\section{Introduction}

  The transition state theory (TST) \cite{Marcelin1915, Eyring1935, Wigner1937} is
  a standard framework for explaining and predicting the reactivity and rate of
  a chemical reaction. The TST is based on two fundamental assumptions:
  (1) All reactive trajectories pass a surface dividing molecular states into reactant
      and product states only once during a chemical reaction process (non-recrossing assumption).
  (2) Reactant molecules are equilibrated in a canonical or microcanonical ensemble
      (quasi-equilibrium assumption).
  However, many experimental and theoretical studies \cite{Carpenter1995, Carpenter1996, Reyes1998, Reyes2000, Nummela2002, Litovitz2008, Goldman2011, Collins2013, Collins2014, Kramer2015, Carpenter2016, Hare2019, Hase1994, Sun2002, Jayee2020, Singleton2003, Ess2008, Hare2017} 
  have reported chemical reactions with a reactivity which cannot be explained by TST
  due to the breakdown of these assumptions by dynamic effects.

  Dynamical systems theory provides a theoretical framework for elucidating
  chemical reaction mechanisms that TST cannot explain. The dynamical systems approach
  describes molecular dynamics by trajectories in a phase space spanned by atomic position
  and momentum coordinates. All reactive trajectories in the phase space are known to
  pass through the interior of a tubular structure connecting reactant and product regions
  through a rank-one saddle region during a chemical reaction process \cite{ClayMarston1989, DeLeon1989, DeAlmeida1990, DeLeon1991a, Deleon1991b, DeLeon1992a, DeLeon1992b}.
  This tubular structure forms a reactivity boundary, separating the reactive trajectories
  of the associated reaction channel from the others. Reactive trajectories flow regularly
  inside the tube in the vicinity of the saddle region, allowing us to define
  a dividing surface that satisfies the non-recrossing assumption \cite{Li2006, Wiggins2001, Ezra2018}.
  Furthermore, the network of reaction tubes corresponding to different reaction
  channels provides essential information for elucidating chemical reaction mechanisms
  that violate the quasi-equilibrium assumption of TST \cite{Li2005, Collins2013, Collins2014, Katsanikas2022}.
  The tube connectivity can be captured by \textit{reactive island} structures---the cross-sections
  of tubes and their interiors on a surface in the phase space (see Sec.~\ref{sec:RI} for details).

  Several algorithms have been developed for computing phase space structures in
  chemical reaction dynamics \cite{Kawai2011a, Mancho2013, Nagahata2020, Naik2021, Mizuno2021}.
  However, these algorithms have been applied
  only to model systems with few degrees-of-freedom (DoF) and their efficiency
  likely deteriorates for real systems with many DoF due to the curse of dimensionality.
  Therefore, dimensionality reduction techniques for phase space structures are indispensable
  to practicalize the theoretical framework for elucidating chemical reaction mechanics
  based on phase space geometry.

  Dimensionality reduction techniques have been already applied to
  chemical reaction dynamics studies, such as ReSPer \cite{Tsutsumi2018, Tsutsumi2020, Tsutsumi2021}
  and PathReducer \cite{Hare2019}.
  These existing methods aim to determine low-dimensional representations of molecular structures
  reflecting structural changes during chemical reactions as much as possible.
  Such dimensionality reduction tasks are formulated as \textit{unsupervised}
  dimensionality reduction in \textit{configuration} space.
  However, dimensionality reduction preserving the information of molecular structural
  changes is not necessarily appropriate for phase space geometry. The information of
  reactivity of each trajectory should be more essential than that of molecular structures
  in dimensionality reduction of reaction tubes (i.e., reactivity boundaries).
  Thus, it is necessary to develop a novel method of determining low-dimensional
  representations of phase space structures preserving the reactivity information
  as much as possible. This task can be formulated as \textit{supervised} dimensionality
  reduction \cite{Chao2019} in \textit{phase} space, where supervised data is
  the reactivity of each trajectory (See Sec.~\ref{sec:dim-reduce-problem} for details).

  In this paper, we propose a method of projecting high-dimensional reactive island structures
  onto low-dimensional phase space using supervised principal component analysis \cite{Barshan2011}.
  The remainder of this paper is organized as follows.
  Sec.~\ref{sec:problem} introduces the reactive island theory and formulates
  the dimensionality reduction problem in phase space geometry.
  Sec.~\ref{sec:algorithm} details our proposed algorithm.
  Sec.~\ref{sec:result} presents numerical experiments and performance evaluation
  of the proposed algorithm applied to a model Hamiltonian system as a proof-of-concept.
  Finally, the paper is concluded in Sec.~\ref{sec:conclusion}.

\section{Problem Formulation} \label{sec:problem}

  \subsection{Reactive island theory} \label{sec:RI}

    Dynamical systems theory of chemical reaction dynamics is based on the phase space
    picture of molecular dynamics. Let us consider $N$-DoF molecular dynamics in gas phase
    governed by a Hamiltonian $H$ with potential energy $V$. Molecular states are represented
    by the atomic positions and momenta of the molecular system, denoted by $(\bm{q}, \bm{p})$.
    Here, $\bm{q}$ and $\bm{p}$ denote the position vector $(q_1, q_2, \dots, q_N)^\top$ and
    the momentum vector $(p_1, p_2, \dots, p_N)^\top$, respectively. Each state is represented by
    a point in the $2N$-dimensional phase space spanned by $\bm{q}$- and $\bm{p}$-axes.
    Dynamical processes are described by trajectories in the phase space, which never
    intersect each other. Under an energy conservation condition $H=E\ (\mathrm{const.})$,
    the energetically accessible region is a $(2N-1)$-dimensional subspace.

    \begin{figure}[!t]
      \centering
      \includegraphics[width=\linewidth]{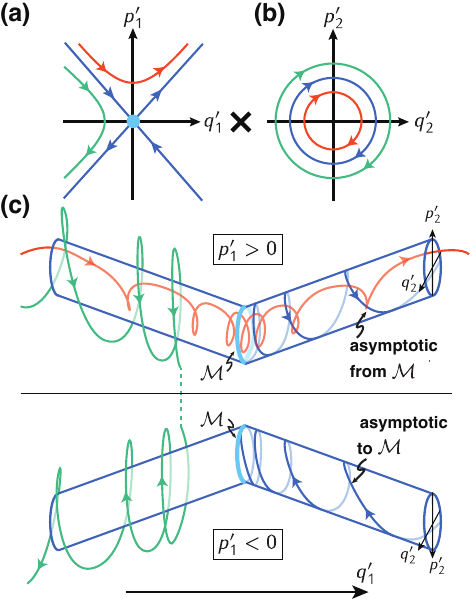}
      \caption{
          Conceptual illustration of \textit{reaction tubes}. We consider a 2-DoF Hamiltonian system,
          though the present discussion can be generalized to any-DoF systems \cite{Nagahata2021}.
          Near a saddle region, we suppose that the 2-DoF is separated into (a) a reactive mode
          $q_1'$ and (b) a thermal bath mode (harmonic oscillator mode) $q_2'$. Panels (a) and (b)
          depict trajectories in phase space of the reactive mode and the bath mode, respectively.
          In panel (a), the red trajectory has sufficient energy in the reactive mode and passes
          through the saddle region, while the green trajectory has insufficient energy and goes back
          to the left region. The four blue trajectories take middle positions between them,
          separating reactive and non-reactive trajectories in the phase space. Given that the total
          energy is constant, the red trajectory in (a) with large mode energy corresponds to
          the red trajectory in (b) with small mode energy. Likewise, the green and blue trajectories
          in (a) also correspond to the trajectories of the same color in (b). Panel (c) schematically
          depicts the phase space of the 2-DoF system. The blue trajectories exhibit spiral motion
          consisting of the one-directional motion in (a) and the circular motion in (b). These spiral
          blue trajectories form tubular structures called \textit{reaction tubes}, separating reactive
          (red) and non-reactive (green) trajectories. Note that all reactive trajectories pass
          through the interior of these tubular structures because they have less energy in the bath
          mode and lie inside the blue circle in (b). In addition, the light blue origin in (a) and
          the circular motion in (b) result in an unstable periodic orbit denoted by $\mathcal{M}$
          in (c). See also Ref.~\cite{Nagahata2021}.
      }
      \label{fig:explain_tube}
    \end{figure}

    One of fundamental phase space structures for elucidating chemical reactivity is
    a \textit{reaction tube}. Figure~\ref{fig:explain_tube} illustrates the concept
    of reaction tubes. A reaction tubes is a $(2N-2)$-dimensional cylindrical structure
    in the phase space, connecting reactant and product regions through a transition
    state region (i.e., a saddle region of $V$). As illustrated in Fig.~\ref{fig:explain_tube},
    a reaction tube forms a \textit{reactivity boundary} that separates reactive and non-reactive
    trajectories in the $(2N-1)$-dimensional energetically-accessible subspace; all reactive trajectories
    pass through the interior of the reaction tube without exception.

\begin{figure}[!t]
      \centering
      \includegraphics[width=\linewidth]{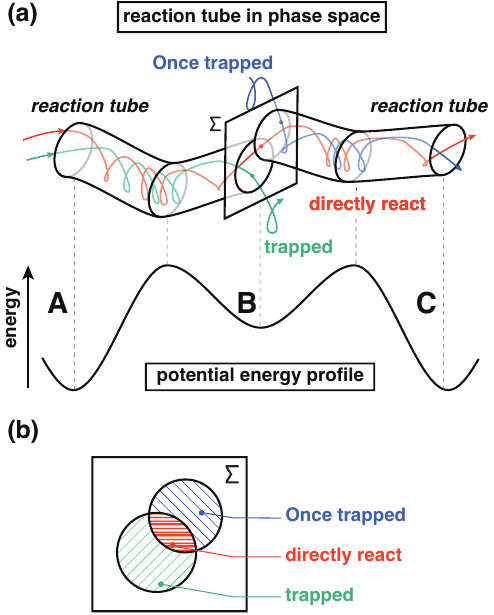}
      \caption{
        Conceptual illustration of \textit{reactive islands}. Here, we consider a 2-DoF
        Hamiltonian system with the potential energy profile shown in panel (a).
        The potential energy surface has three local minima A, B, and C, corresponding to
        reactant, intermediate, and product, respectively. In panel (a), reaction tubes of
        reactions A $\to$ B and B $\to$ C are depicted. The surface designated by $\Sigma$ is
        a two-dimensional surface at the local minimum B. The surface $\Sigma$ cuts
        the reaction tubes and the surface of section is shown in panel (b).
        The intersection of a reaction tube with $\Sigma$ forms a closed curve.
        The inside of a closed curve consists of intersections of all reactive trajectories
        with $\Sigma$. This region of the intersections of reactive trajectories enclosed by
        that of reaction tubes is called \textit{reactive island}. In the present example,
        the reactive islands  of reactions A $\to$ B and B $\to$ C have overlap. The red
        trajectory in panel (a) passes the overlap region, indicating that it directly reacts
        A $\to$ C without trapped in the intermediate region B. In contrast, the green/blue
        trajectories passing the green/blue reactive island regions in panel (b) are
        reactive trajectories with trapped in B after/before reactions, respectively.
        This classification of reactive trajectories is neglected in TST due to
        the quasi-equilibrium assumption.
      }
      \label{fig:explain_ri}
    \end{figure}

    \textit{Reactive islands} provide essential dynamical information to elucidate a reaction system
    with multiple reaction channels. Figure~\ref{fig:explain_ri} illustrates the concept
    of reactive islands. In the figure, we consider a reaction system with two channels
    A $\to$ B and B $\to$ C. Panel (b) shows the cross sections of the two reaction tubes
    corresponding to the two reaction channels on a surface $\Sigma$ in panel (a).
    The cross section of the reaction tubes and their interiors form island-like patterns
    on $\Sigma$, called \textit{reactive islands} (RIs). In the present case, the two RIs overlap.
    This indicates that there exist reactive trajectories directly reacting from
    A to C without trapped in the intermediate region B. The existence of such bypassing
    reactive trajectories is known as \textit{dynamic matching} \cite{Carpenter1995} and
    neglected in TST due to the quasi-equilibrium assumption. In this way, RIs
    are crucial for chemical reaction analyses taking into account dynamic effects.

    As another example, let us consider a reaction system with three channels,
    (i)A $\to$ B, (ii)B $\to$ C, and (iii)B $\to$ D. Considering a cross-section
    $\Sigma$ in the B's well, there should be three RIs corresponds to
    the three reaction channels. If the RIs of (i) and (ii) overlap,
    it indicates that there is the direct reaction path A $\to$ C.
    Similarly, the overlap of the RIs of (i) and (iii) indicates the existence of
    the direct reaction A $\to$ D. The selectivity between two products C and D
    may be estimated by these RI overlaps\cite{DeLeon1991a}.

  \subsection{Dimensionality reduction problem} \label{sec:dim-reduce-problem}
    RIs are $(2N-2)$-dimensional structures for an $N$-DoF system; thus a low-dimensional
    projection of reactive islands is necessary for practical analysis of chemical reaction
    dynamics with many-DoF. In this subsection, we present a general problem formulation of
    the dimensionality reduction of RIs. A prototypical algorithm for this problem will be
    proposed in the next section.

    The dimensionality reduction of RIs aims to determine a mapping $f$
    from a $(2N-2)$-dimensional surface $\Sigma\ (\subset \mathbb{R}^{2N})$
    to a $d$-dimensional space preserving the RI structure as much as possible.
    Here, $d$ is a user-specified integer number satisfying $d < 2N-2$.
    Note that how to determine $d$ is also a critical issue as well as how to compute $f$.
    An appropriate protocol of determining $d$ may vary depending on a dimensionality reduction
    algorithm and a reaction system of interest. We provide a specific protocol for our proposed
    algorithm in Sec.~\ref{sec:algorithm} and a further perspective in Sec.~\ref{sec:conclusion}.
    In the remainder of this subsection, we specify the structure of $f$ and measures of
    the preservation of the RI structure. 

    We construct the mapping $f$ by composition of a coordinate transformation $\phi$
    in the $2N$-dimensional phase space and a projection $\pi$ onto the $d$-dimensional space.
    Here, the projection $\pi$ is defined as
    $\pi(z_1, z_2, \dots, z_d, \dots, z_{2N})=(z_1, z_2, \dots, z_d)$.
    Restricting the domain of $\phi$ to $\Sigma$, we define $f=\pi\circ\phi|_\Sigma$.
    For instance, a dimensionality reduction that projects molecular states onto
    a subspace spanned by $r$ dominant normal modes can be constructed by
    the normal mode transformation $\phi$ and a standard projection
    $\pi$ onto the $2r$-dimensional subspace.

    We measure the degree of the RI structure preservation by the dependence between
    the coordinate variable in the $d$-dimensional subspace and its corresponding
    reactivity label variable. The reactivity label is, for example, the colored pattern
    in Fig.~\ref{fig:explain_ri}(b). According to the RI theory, the reactivity label is
    a dependent variable of the coordinate variable. A dimensionality reduction, however,
    may map points of different reactivity labels to a single point on the $d$-dimensional
    subspace. This mixing of reactivity labels decreases the degree of the dependence between
    the coordinate variable and the reactivity label variable in the $d$-dimensional subspace.
    In other words, we define the dependence as the relationship that specifying the coordinates
    determines the reactivity label. This dependence should be conserved as much as possible
    during a dimensionality reduction as the separation of reactivity is the essence of
    the concept of RIs.

    Therefore, we formulate the dimensionality reduction problem as the following
    variational problem:
    \begin{equation}
        \label{eq:max_d}
        \begin{aligned}
          &\operatorname*{maximize}_{\phi\colon\mathbb{R}^{2N}\to\mathbb{R}^{2N}}
          \quad &&\mathcal{D}\left(\bar{\bm{\mathcal{Z}}};\ \mathcal{Y} \right) \\
          &\text{subject to} &&
          \bar{\bm{\mathcal{Z}}} = \pi\circ\phi|_\Sigma(\bm{\mathcal{Z}})
        \end{aligned}
    \end{equation}
    where $\mathcal{D}(\cdot ; \cdot)$ denotes a measure of the dependence of two random variables,
    $\bm{\mathcal{Z}}$ and $\bar{\bm{\mathcal{Z}}}$ signify the coordinate variables
    in the original phase space and the $d$-dimensional subspace, respectively, and
    $\mathcal{Y}$ is the reactivity label variable. Several dependence measures can be
    employed as $\mathcal{D}$, including the Hilbert--Schmidt independence criterion
    \cite{Gretton2005} and the mutual information. Different classes of the coordinate transformation
    $\phi$ and different dependence measures $\mathcal{D}$ derive different algorithms for
    the dimensionality reduction of RIs. The next section presents one of them as
    a prototypical algorithm.

\section{Proposed Algorithm} \label{sec:algorithm}

  This section presents an algorithm for solving the dimensionality reduction problem
  given by Eq.~\eqref{eq:max_d} using the supervised principal component analysis
  (SPCA) \cite{Barshan2011}. SPCA employs the Hilbert--Schmidt independence criterion (HSIC)
  \cite{Gretton2005} as the dependence measure $\mathcal{D}$ and restricts the coordinate
  transformation $\phi$ to orthogonal transformation. The HSIC measures the dependence
  between two random variables by the Hilbert--Schmidt norm of the cross-covariance
  operator---the larger HSIC is, the higher the dependence is. For the detailed background
  of HSIC and SPCA, see the original paper \cite{Barshan2011}.

  The SPCA algorithm requires a data matrix $\bm{Z}$ and a label matrix $\bm{Y}$ as input.
  In the present case, the data matrix is a $2N \times m$ matrix whose columns are
  the $2N$-dimensional coordinate vectors of $m$ points on a surface $\Sigma$.
  As a preprocessing, all parameters were scaled such that their minimum value is 0 and the maximum value is 1.
  Then, in the process of SPCA, we assume that the data matrix is centered in advance, that is, the origin of
  the coordinate vectors is set to their mean.
  The label matrix is a $1 \times m$ matrix where the $m$-th column corresponds to
  the reactivity label associated with the $m$-th column of $\bm{Z}$. Our proposed algorithm
  supposes that the reactivity label can be determined using numerical simulation of
  trajectories initiated from the $m$ points on $\Sigma$. The labeling method may vary
  depending on a reaction system of interest (see Sec.~\ref{subsec:modelHamiltonian} for
  a specific example).

  The orthogonal coordinate transformation $\phi$ and standard projection $\pi$ can be represented by
  a $2N \times 2N$ orthogonal matrix $\bm{U}$ and a $d \times 2N$ projection matrix $\bm{P}$.
  Here, the elements of row $i$ and column $j$ of $\bm{P}$ is the Kronecker delta $\delta_{ij}$.
  The projected data matrix is written as
  \begin{equation}
    \bar{\bm{Z}} = \bm{P} \bm{U}^\top \bm{Z}.
    \label{eq:projected-data-matrix}
  \end{equation}
  For later convenience, we use $\bm{U}^\top$ in Eq.~\eqref{eq:projected-data-matrix}
  instead of $\bm{U}$.

  The HSIC between the projected data and label is empirically estimated from the data
  as follows: 
  \begin{equation}
    \mathrm{HSIC} = \frac{1}{(2N - 1)^2} \mathrm{tr} (\bm{KL}).
  \end{equation}
  Here, $\operatorname{tr}$ denotes the matrix trace. The matrix $\bm{K}$ is a data kernel
  which is written in the present formulation as
  \begin{equation}
    \bm{K} = \bar{\bm{Z}}^\top \bar{\bm{Z}}
           = \bm{Z}^\top \bm{U} \bm{P}^\top \bm{P} \bm{U}^\top \bm{Z}.
  \end{equation}
  The matrix $\bm{L}$ is a label kernel calculated from the label matrix $\bm{Y}$.
  In this paper, the entry $l_{i,j}$ is set to 1 if $i$-th data and $j$-th data have
  the same label, otherwise 0.
  
  The variational problem of Eq.~\eqref{eq:max_d} is now given by
  \begin{equation}
    \begin{aligned}
      & \operatorname*{maximize}_{\bm{U}\in \mathbb{R}^{2N\times2N}} \quad
      &&\operatorname{tr}(\bm{Z}^\top \bm{U} \bm{P}^\top
                      \bm{P} \bm{U}^\top \bm{Z} \bm{L}) \\
      & \text{subject to} && \bm{U}^\top\bm{U} = \bm{I}.
    \end{aligned}
  \end{equation}
  Note that the matrix trace to be maximized is rewritten as
  \begin{equation}
    \begin{split}
    &\operatorname{tr}(\bm{Z}^\top \bm{U} \bm{P}^\top
                      \bm{P} \bm{U}^\top \bm{Z} \bm{L}) \\
    =
    &\operatorname{tr}\left(\bm{P} (\bm{U}^\top \bm{Z} \bm{L}
                            \bm{Z}^\top \bm{U}) \bm{P}^\top\right).
    \end{split}
  \end{equation}
  Since matrix $\bm{P}$ is a projection matrix, this matrix trace equals
  to the sum of the first $d$ diagonal elements of matrix $\bm{U}^\top \bm{ZLZ}^\top \bm{U}$.
  Given that $\bm{ZLZ}^\top$ is symmetric and $\bm{U}$ is an orthogonal matrix,
  this matrix trace takes the maximum value when the first $d$ columns of $\bm{U}$ are
  the eigenvectors associated with the top $d$ eigenvalues of $\bm{ZLZ}^\top$.
  Therefore, this optimization problem is reduced to the eigenvalue problem of
  matrix $\bm{ZLZ}^\top$. This reduction is similar to the original (unsupervised) PCA
  where the empirical sample covariance matrix $\bm{Z}\bm{Z}^\top$ is diagonalized
  through the eigenvalue problem. Indeed, when we set the label kernel $\bm{L}$ to
  the identity matrix, the SPCA is equivalent to the original PCA. Finally, we summarize
  the whole procedure of our proposed algorithm as a pseudo code in Fig.~\ref{alg:alg_spca_ri}.

  As we mentioned above, how to determine the dimensionality $d$ is one of the important issues.
  In our framework, we determine $d$ based on the cumulative contribution rate---the ratio
  of the sum of top $d$ eigenvalues to the sum of all eigenvalues of matrix $\bm{ZLZ}^\top$.
  The larger this value is, the more the dimensionality reduction preserves the dependence
  between the projected coordinates and its reactivity label. An appropriate value of $d$
  can be determined such that the cumulative contribution rate for $d$ is larger than
  a specified threshold (close to one).

  \begin{figure}[!t]
    \begin{algorithm}[H]
    \caption{Compute a low-dimensional projection of reactive islands}
    \begin{algorithmic}[1]
        \REQUIRE $d$---The diminsionality of a projected space,
                 $m$---The number of samples of initial conditions on a surface $\Sigma$.
        \ENSURE $\bar{\bm{Z}}$---Phase space coordinates of $m$ points in the projected space,
                $\bm{Y}$---Reactivity labels of the $m$ points.
        \FOR{$i=1$ \textbf{to} $m$}
          \STATE $\bm{z}_i \gets$ An initial condition sampled on $\Sigma$
          \STATE $trj_i \gets$ A calculated trajectory initiated from $\bm{z}_i$
          \STATE $y_i \gets$ The reactivity label of $trj_i$
        \ENDFOR
        \STATE $\bm{z}_i$ is scaled to the range 0 to 1 (min-max scaling)
        \STATE $\bm{z}_\mathrm{mean} \gets m^{-1}\sum_{i=1}^m \bm{z}_i$
        \STATE $\bm{Z} \gets [\bm{z}_1-\bm{z}_\mathrm{mean}, \dots, \bm{z}_m-\bm{z}_\mathrm{mean}]$
        \STATE $\bm{Y} \gets [y_1, \dots, y_m]$
        \STATE $\bm{L} \gets [\delta_{y_i, y_j}]_{1 \le i \le m, 1 \le j \le m}$
        \STATE $\bm{Q} \leftarrow \bm{ZLZ}^\top$
        \STATE $\bm{U} \gets$ The matrix of the eigenvectors of $\bm{Q}$
        \STATE $\bm{P} \gets$ The projection matrix from $2N$ to $d$ dimension
        \STATE $\bar{\bm{Z}} \leftarrow \bm{P} \bm{U}^\top \bm{Z}$
        \RETURN $\bar{\bm{Z}}, \bm{Y}$
    \end{algorithmic}
    \end{algorithm}
    \caption{Pseudo code of the proposed algorithm.
             This algorithm depends on (1) an initial condition sampler from a surface $\Sigma$,
             (2) a numerical integration method for the trajectory calculation,
             (3) a reactivity labeling method, and (4) an eigensolver for $\bm{Q}$,
             as external subroutines.}
    \label{alg:alg_spca_ri}
  \end{figure}

\section{Numerical demonstration and performance evaluation}
  \label{sec:result}

  This section presents a numerical demonstration of the proposed algorithm.
  We first define model Hamiltonian systems and numerical procedure in Sec.~\ref{subsec:modelHamiltonian}.
  Next, we show projected RIs of the model systems computed by the proposed algorithm in Sec.~\ref{sec:numerical-results}.
  Finally, the performance of the proposed algorithm is evaluated from two different perspective:
  (1) the quality of reactivity label prediction (Sec.~\ref{sec:label-prediction}) and
  (2) the clearness of reactivity boundary (Sec.~\ref{sec:boundary-detecton}).

  \subsection{Model Hamiltonian system and numerical procedure}
    \label{subsec:modelHamiltonian}

\begin{figure}
        \centering
        \includegraphics[width=\linewidth]{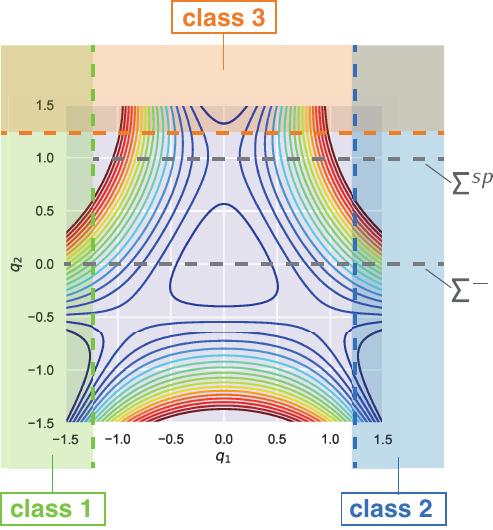}
        \caption{A contour plot of the Hénon-Heiles potential energy landscape. 
        The green, blue, and orange regions are product regions, classes 1, 2, and 3,
        respectively. The gray dotted lines denote lines (hyperplanes for high-dimensional systems)
        $q_2=0$ ($\Sigma^-$) and $q_2=1$ ($\Sigma^\mathrm{sp}$). In the present numerical
        demonstration, RIs on the surface $\Sigma^-$ were computed, whereas trajectory calculations
        were started from $\Sigma^\mathrm{sp}$. See the text for the detailed calculation procedure.
        }
        \label{fig:henon-heiles_potential}
      \end{figure}

      The model Hamiltonian system that we used is based on the Hénon-Heiles system\cite{Henon1964}.
      The system is a two-DoF Hamiltonian system with one potential minimum and three index-one saddles
      which define three exit channels to ``products.'' The 2-DoF Hénon-Heiles Hamiltonian $H_2$ is
      given by
      \begin{multline}
        H_2(q_1, q_2, p_1, p_2)\\
        = \frac{p_1^2}{2} + \frac{p_2^2}{2} 
          + \frac{q_1^2}{2} + \frac{q_2^2}{2}
          + q_1^2 q_2 - \frac{1}{3}q_2^3.
      \end{multline}
      The potential energy surface of this system is shown in Fig.~\ref{fig:henon-heiles_potential}.

      We extend the two-DoF Hénon-Heiles system to $N$-DoF systems ($N > 3$) by adding
      vibrational bath modes to the original system.
      The extended $N$-DoF Hénon-Heiles Hamiltonian $H_N$ is defined as
      \begin{equation}
          H_N(\mathbf{q}, \mathbf{p}) 
            = H_2 + \sum_{i=3}^N \frac{1}{2} (p_i^2 + \omega_i^2 q_i^2 +  c_i q_2 q_i^2).
      \end{equation}
      In this paper, we set all frequencies $\omega_i\ (i=3, \dots, N)$ to one.
      All bath modes are coupled with the original $q_2$ coordinate and each coupling constant
      $c_i\ (i=3, \dots, N)$ is set to $c_i=(-1)^{(i+1)}/2$. This extended $N$-DoF system is designed
      so that it has one potential energy minimum and three saddle points at the same positions
      as the original system and that there are no additional equilibrium points in the range of
      $|q_i| \le 2\ (i=1,\dots,N)$ for $N \le 10$.

       The present numerical demonstration aims to compute RIs on a surface $\Sigma^-$
       defined as
      \begin{equation}
        \Sigma^- = \left\{ (\mathbf{q}, \mathbf{p}) | q_2 = 0, p_2 < 0, H(\mathbf{q}, \mathbf{p}) = E \right\}.
      \end{equation}
      This surface traverses the potential minimum $\bm{q}=0$. In particular, we focus on
      the overlaps between the reaction tube entering the basin around $\bm{q}=0$ through
      the upper saddle region and those exiting from the basin through the three exit channels.
      In other words, the present calculation aims to examine whether there exist direct reactive
      trajectories from class 3 to class 1, 2, or 3 through the intermediate basin region.

      To sample data points inside the entering reaction tube efficiently, we sample
      trajectories initiated from a surface $\Sigma^{\mathrm{sp}}$ defined as
      \begin{equation}
        \Sigma^{\mathrm{sp}} =
        \left\{ (\mathbf{q}, \mathbf{p}) \mid
               q_2 = 1, p_2 < 0, H(\mathbf{q}, \mathbf{p}) = E \right\}.
      \end{equation}
      This surface traverses the upper saddle point; thus the sampled trajectories are considered
      as reactive trajectories inside the entering reaction tube. At the saddle point,
      the reactive mode is mode 2, which is decoupled from the other non-reactive modes.
      Therefore, we first sample initial conditions for the non-reactive modes
      $(q_1, q_3, ..., q_N, p_1, p_3, ..., p_N)$ at uniformly random
      such that the following condition is met:
      \begin{equation}
          \frac{1}{2}\left(p_1^2 + 3q_1^2\right)
          + \sum_{i=3}^N \frac{1}{2}\left(p_i^2 + (1+c_i)q_i^2\right)
          \leq E - \frac{1}{6}.
      \end{equation}
      Here, the left and right hand sides correspond to the Hamiltonian of the non-reactive subsystem
      and the excess energy at the saddle point, respectively. Next, we calculate
      initial conditions for the reactive mode momentum $p_2$ as
      \begin{equation}
          p_2 = - \sqrt{
                     2E - \frac{1}{3} -
                     \left\{ p_1^2 + 3q_1^2 + 
                             \sum_{i=3}^N \left(p_i^2 + (1+c_i)q_i^2\right) \right\}
                  }.
      \end{equation}
      Note that $q_2$ is always one on $\Sigma^{\mathrm{sp}}$. Finally, every intersection of
      a sampled trajectory with $\Sigma^-$ is used as a data point $\bm{z}_i$.

      The reactivity label of each trajectory is determined by the following criterion:
      A sampled trajectory is labeled as class 1, 2, or 3 if it passes $\Sigma^-$ only once
      and satisfies one of the following conditions within a maximum simulation time
      $t_{\mathrm{term}}$:
      \begin{description}
        \item[class 1] $q_1 < -1.25$ (the left green region in Fig. \ref{fig:henon-heiles_potential})
        \item[class 2] $q_1 > 1.25$ (the right blue region in Fig. \ref{fig:henon-heiles_potential})
        \item[class 3] $q_2 > 1.25$ (the upper orange region in Fig. \ref{fig:henon-heiles_potential})
      \end{description}
      As we focus on direct reactive trajectories, we label a trajectory as \textbf{class 0},
      indicating ``not directly reactive'', if the trajectory passes $\Sigma^-$ more than once
      or the simulation time reaches $t_{\mathrm{term}}$ before the above three conditions are met.
      That is, all intersection points along ``not direct reactive'' trajectories are labeled as \textbf{class 0}.
      In this paper, we set $t_{\mathrm{term}}=50$. This value is about eight times longer than
      the one period of the Hénon-Heiles harmonic oscillator mode $T = 2 \pi$.
      Thus the simulation time is considered to be long enough to distinguish direct reactive
      trajectories from others.

      Numerical simulations were conducted for the model systems of 4, 6, 8, and 10 DoFs.
      The total energy was set so that the average energy for each DoF was 0.1. 
      Therefore, for 4, 6, 8, and 10 DoF systems, the total energy $E$ was set to 0.4, 0.6, 0.8,
      and 1.0, respectively. A Python program we implemented performed the whole protocol
      presented above. The trajectory calculation was performed with \textsf{scipy.integrate.odeint}
      with time step 0.01.

  \subsection{Numerical results of the proposed algorithm}
      \label{sec:numerical-results}

\begin{figure*}
        \centering
        \includegraphics[width=\linewidth]{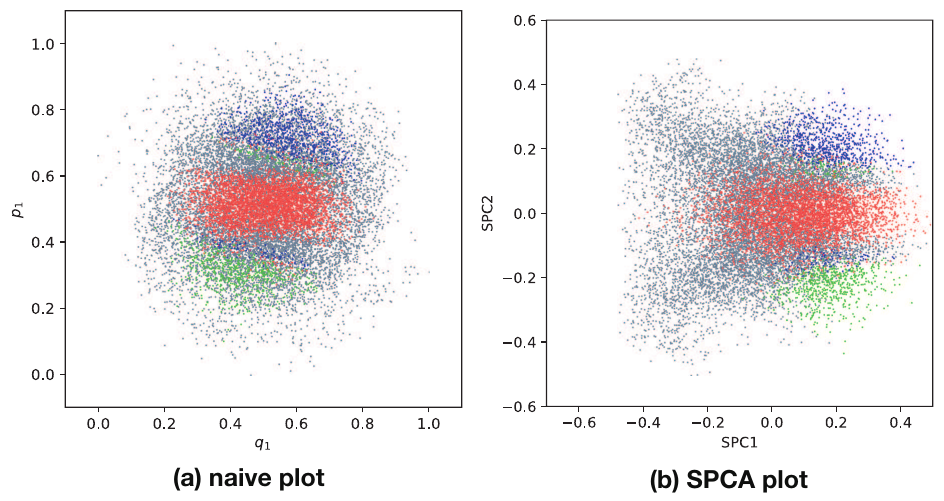}
        \caption{
          Projected RIs on two-dimensional planes for the 8-DoF Hénon-Heiles system.\footnote{The total number of the data is 17073, and all data are projected in this figure.}
          (a) ``naive'' projection---data points are projected onto the plane spanned by
          $q_1$ and $p_1$.
          (b) our proposed algorithm---data points are projected onto the plane
          spanned by the top two principal components computed by the SPCA.
          The color of each data point indicates the reactivity label of
          the point---class 0: gray, class 1: green, class 2: blue, and class 3: red.
        }
        \label{fig:result_of_projection}
      \end{figure*}

\begin{figure}
        \centering
        \includegraphics[width=\linewidth]{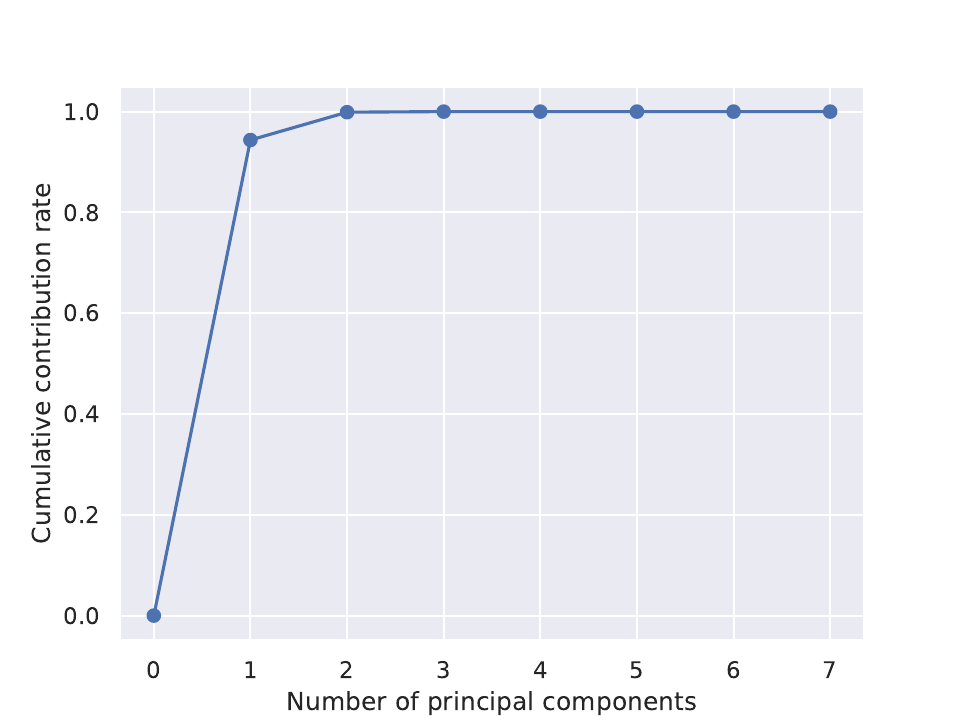}
        \caption{
          Cumulative contribution rates of the SPCA result for the 8-DoF system.
          This figure shows the cumulative contribution rate up to the top seven
          principal components. The top three eigenvalues are $9.4 \times 10^{-1}$,
          $5.6 \times 10^{-2}$, $1.1 \times 10^{-3}$, while the remaining eigenvalues
          are less than $10^{-16}$.
        }
        \label{fig:cum_cont_rate}
      \end{figure}

      Figure~\ref{fig:result_of_projection} shows numerical results for the 8-DoF system.
      Based on the cumulative contribution rate shown in Fig.~\ref{fig:cum_cont_rate},
      we set the dimensionality of the projected subspace $d$ to two for this case.
      Panels (a) depicts the result of ``naive'' projection onto the first reactive mode
      of the original Hénon-Heiles system, and panel (b) displays the results of our proposed
      algorithm. Here, the naive projection result is shown as a reference for comparison
      and performance evaluation describe below. While the axes of the naive projection result
      are the first reactive mode $(q_1, p_1)$, those of the SPCA result are superpositions of
      all reactive and non-reactive modes. Table \ref{tab:construction_pc} shows the coefficients
      of the original position and momentum variables in the top-three principal components (PCs)
      computed by the SPCA. Additionally, we present results for the 4, 6, and 10-DoF systems in
      Supplementary Material.

      Compared to the RIs projected by the naive projection [Fig.~\ref{fig:result_of_projection}~(a)],
      those projected by our SPCA algorithm [Fig.~\ref{fig:result_of_projection}~(b)] appear to be
      less mixed with other regions; especially, the rightmost part of each projected RI consists of
      points with the same reactivity. These results indicate the superiority of our SPCA algorithm
      over the naive projection. In what follows, we will assess the superiority of SPCA
      in objective manners.

      \begin{table}
        \caption{Compositions of the top-three principal components (PCs) for the 8-DoF system.
        }
        \label{tab:construction_pc}
        \centering
        \begin{tabular}{lrrrr}
          \hline
               &       $q_1$  &      $p_1$      &      $q_2$      &   $p_2$ \\
          \hline
          PC1 &  -0.014 & -0.0026 &  0.0 & 0.99 \\
          PC2 &  -0.34 &  -0.93 & 0.0 & -0.0060 \\
          PC3 &   0.27 & -0.080 & 0.0 & 0.044 \\
          \hline \hline
              &        $q_3$  &      $p_3$   &     $q_4$      &   $p_4$ \\
          \hline
          PC1 &  -0.0050 & -0.014 & -0.025 & 0.015  \\
          PC2 &   0.022 & 0.028 & 0.017 & -0.014 \\
          PC3 &   0.23 & 0.13 & 0.20 & -0.28 \\
          \hline \hline 
             &         $q_5$ &        $p_5$ &        $q_6$ &        $p_6$ \\
          \hline
          PC1 &  -0.0077 & 0.0012 & -0.017 &  0.013 \\
          PC2 &   0.010 & -0.0022 & 0.010 & -0.021 \\
          PC3 &  0.088 & -0.38 & 0.42 & -0.24 \\
          \hline \hline
             &          $q_7$ &       $p_7$ &        $q_8$ &        $p_8$   \\
          \hline
          PC1 &   0.022 & 0.0098 &  0.013 & -0.016    \\
          PC2 &   0.0058 & 0.018 & 0.016 & 0.015    \\
          PC3 &  -0.42 & -0.35 & -0.076 &  0.12   \\
          \hline  
        \end{tabular}
      \end{table}

      Table~\ref{tab:HSIC_ratio} summarizes the ratio of the empirical HSIC of the SPCA result
      to that of the naive projection result for each model system. As the empirical HSIC is
      a measure of the dependence between positions on a low-dimensional subspace and reactivity labels,
      those ratios indicate how much the proposed algorithm preserve the dependence between positions
      and reactivity labels compared with the naive projection. The table shows that all the ratios are
      sufficiently large, specifically ranging from about 15 times to 90 times.
      These large ratios indicate that the proposed algorithm is more effective compared with
      the naive projection as a dimensionality reduction method for RI calculation.
      In addition, according to the table, the ratio decreases as $N$ increases.
      However, this tendency may be improved by more advanced method such as
      nonlinear kernel dimensionality reduction method. Applying such advanced methods
      is beyond the scope of this paper and remains as future work.

      \begin{table}
        \caption{
          The ratios of the empirical HSICs of the SPCA results to those of the naive projection results
          for 4, 6, 8, and 10-DoF systems. The ratio value greater than one indicates the SPCA algorithm
          preserves more information on the dependence between positions and reactivity labels
          compared to the naive dimensionality reduction.
        }
        \label{tab:HSIC_ratio}
        \centering
        \begin{tabular}{lcccc}
          \hline
            & 4 DoF & 6 DoF & 8 DoF & 10 DoF \\
          \hline
          HSIC(SPCA/naive) & 95  & 24  & 18 & 13 \\
          \hline  
        \end{tabular}
      \end{table}

  \subsection{Effectiveness evaluation 1: label prediction}
    \label{sec:label-prediction}

    An RI pattern provides a predictor of the reactivity label from the phase space coordinate.
    In particular, the predictor of RIs without dimensionality reduction is a perfect predictor;
    once we determine an initial condition, we can predict the reactivity of the trajectory starting
    from the initial condition with 100\% accuracy due to the deterministic property of Hamiltonian dynamics.
    Dimentionality reduction of RIs decreases the prediction accuracy in general, but the degree of
    the accuracy decrease depends on the dimensionality reduction method. Thus, the quality of the reactivity
    label prediction is one of the effectiveness indicators of dimensionality reduction methods for RI calculation.

    To evaluate the label prediction accuracy, we constructed the k-nearest neighbor predictors
    \cite{Cover1967} from the projected labeled data points shown in Fig.~\ref{fig:result_of_projection}
    and Supplementary Material. A k-nearest neighbor algorithm predicts a reactivity label of a point
    by majority voting of the labels of the $k$ nearest data points in a low-dimensional space.
    Here, $k$ is a hyperparameter of the predictor.
    To predict the labels, the data was divided into training and test sets such that the test data constituted 10\% of the total.

    Furthermore, we employ the macro-F1 score \cite{Manning2008, Sokolova2009, Takahashi2022} as a prediction performance score,
    which is widely-used for the performance evaluation of multi-class classifiers.
    The macro-F1 score is based on the F1 score \cite{Chinchor1992} for binary classification between
    reactivity class $l$ ($l=0,1,2,3$) and the others. The F1 score for the binary classification
    is defined as
    \begin{align*}
        F_1 = \frac{2 \cdot \mathrm{Recall} \cdot \mathrm{Precision}}{\mathrm{Recall} + \mathrm{Precision}},
    \end{align*}
    where
        \begin{align*}
        \mathrm{Recall} &= \frac{\mathrm{TP}}{\mathrm{TP}+\mathrm{FN}}, \\
        \mathrm{Precision} &= \frac{\mathrm{TP}}{\mathrm{TP}+\mathrm{FP}}.
    \end{align*}
    Symbols $\mathrm{TP}$, $\mathrm{FP}$, and $\mathrm{FN}$ designate the number of prediction instances
    of true positive, false positive, and false negative, respectively. Here, `positive' and `negative'
    indicate a predicted binary label (positive: class $l$, negative: the others), while `true' and `false'
    indicate whether the prediction is correct or not. The macro-F1 score is the arithmetic mean of
    the F1 score for each class:
    \begin{align*}
        \mathrm{macro\mathchar`-}F_1 = \frac{\sum_{l} F_1(\mathrm{class} \; l)}{4}.
    \end{align*}
    This value is a scalar quantity uniquely determined for each result of dimensionality reduction.
    This macro-F1 score is suitable for the performance evaluation of predicting rare reaction events
    (See Supplementary Material for detailed discussion).

    We calculated the ratio of the macro-F1 score for the proposed algorithm to that for the naive projection.
    The macro-F1 score ratio was calculated for the case of $k=3, 5, 7, 9, 11, 13$, and $15$ for each-DoF system.
    The range of the ratio values is between 1.25 and 1.7, thus the macro-F1 ratio is greater than one
    for all cases. This result shows that the predictive performance of the proposed algorithm
    is better than the naive dimensionality reduction, irrespective of the choice of the hyperparemeter $k$.
    For more detailed results, see Supplementary Material.
    
  \subsection{Effectiveness evaluation 2: boundary clearness}
  \label{sec:boundary-detecton}
    RI boundaries are cross-sections of reaction tubes on a phase space surface and specifies a RI pattern.
    The RI boundary detection is an essential part of efficient algorithms for computing RIs \cite{Nagahata2020,Mizuno2021}.
    These algorithms (1) estimate provisional RI boundaries from current data points, (2) sample new points from
    (the vicinity of) the provisional boundaries, and (3) compute the reactivity label of the sampled points
    based on trajectory calculation. This process refines provisional RI boundaries. Repeating 
    the boundary refinement process, we can compute RI boundaries---thus RI patterns---with satisfactory accuracy.

    There are several methods for estimating RI boundaries. The algorithm proposed by Nagahata \textit{et al.}
    estimates RI boundaries by the asymptotic trajectory indicator \cite{Nagahata2020}. This indicator is based on
    a fact that the time required for a trajectory to pass through a saddle region increases and diverges to infinity
    as it approaches a reaction tube. Therefore, long passage time is an indicator of an RI boundary.
    The algorithm proposed by Mizuno \textit{et al.} estimates RI boundaries by Voronoi tessellation of
    labeled data points. This boundary detection method is based on a simple fact that RI boundaries
    separate regions with different reactivity labels.

    The dimensionality reduction of RIs blurs RI boundaries in general. For example, as shown
    in Fig.~\ref{fig:result_of_projection}, different-reactivity points are mixed in some areas
    on the low-dimensional space, making ``fuzzy'' RI boundaries. This fuzziness of projected RI
    boundaries makes the aforementioned boundary refining process less efficient. Thus, the clearness
    of reactivity boundaries is one of the effectiveness indicators of dimensionality reduction methods
    for RI calculation. In the remainder of this subsection, we compare reactivity boundaries
    projected by the proposed algorithm and the naive projection based on the two criteria,
    (1) reactivity label separation (employed by Mizuno \textit{et al.} \cite{Mizuno2021}) and
    (2) trajectory passage time (employed by Nagahata \textit{et al.} \cite{Nagahata2020}).

    \subsubsection{Reactivity label separation}
    \label{sec:subsub_viz_reactive}

        RI regions projected onto a low-dimensional space can be estimated based on a predicted probability
        $p(l|x)$ that the reactivity label is $l$ at point $x$ on the projected space \cite{Mizuno2021}.
        The probability $p(l|x)$ can be viewed as the grade of membership of $x$ in the RI of $l$ as a fuzzy set.

\begin{figure*}
          \centering
          \includegraphics[width=\linewidth]{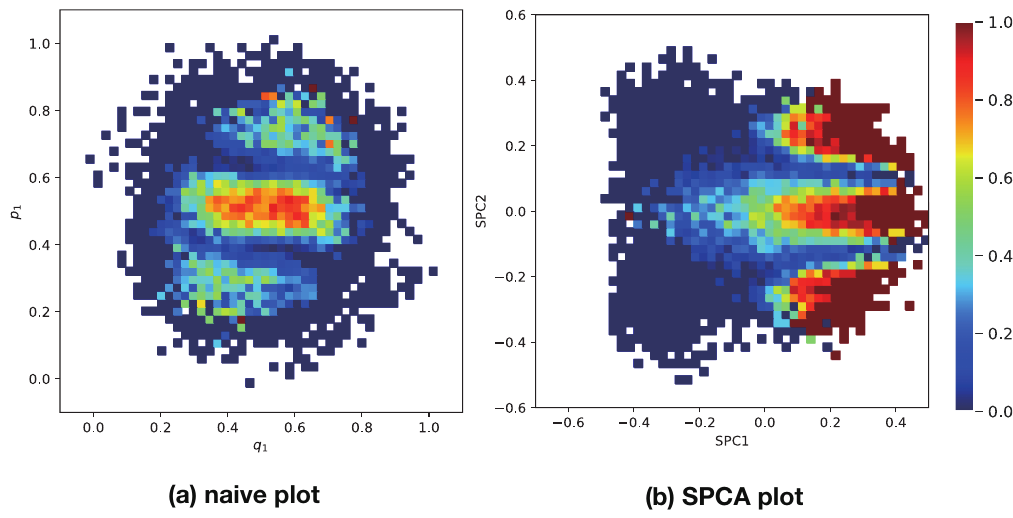}
          \caption{
            Heatmaps of the certainty of RI attribution [Eq.~\eqref{eq:RI-attribution-certainty}]
            for the 8-DoF system. To plot the heatmaps, we divided each low-dimensional space
            into grids and estimated the probability $p(l|x)$ by the ratio of label $l$ among
            the data points in the grid of point $x$.
            This heatmaps are calculated from 17073 data points.
          }
          \label{fig:upper_dec_8dof}
        \end{figure*}

        To visualize the fuzziness of projected RIs, we plotted heatmaps of the following indicator $C$:
        \begin{equation}
          C(x) = \max_{l \in \{1, 2, 3\}} p(l|x).
          \label{eq:RI-attribution-certainty}
        \end{equation}
        This value indicates the certainty (or purity) of RI attribution at point $x$.
        Figure~\ref{fig:upper_dec_8dof} shows the results for the proposed algorithm and the naive projection.
        
        In the SPCA case, red grids with high $C$ value form clusters, each of which corresponds to an RI.
        The surrounding areas of the red clusters with color gradation from red to blue correspond to
        fuzzy RI boundaries. On the other hand, there are few red grids and the color gradation corresponding
        fuzzy RI boundaries---especially for classes 1 and 2---is unclear in the naive projection case.
        As noted in the beginning of this subsection, efficient algorithms \cite{Nagahata2020,Mizuno2021}
        compute RI boundaries by sampling data points in the vicinity of provisional boundaries.
        In other words, these algorithms save the cost of computing data points with high-confidence
        label estimates, i.e., high $C$ values in the present case. Since grids with high $C$ values
        are more clearly separated from fuzzy RI boundary regions in the SPCA case than 
        in the naive projection case, the SPCA-based projection should be more suitable for
        applying the efficient algorithm for RI calculation.
        These results suggest that the proposed algorithm using SPCA is more effective than the naive projection
        in terms of the clearness of fuzzy RI boundaries.

    \subsubsection{Trajectory passage time}
    \label{sec:subsub_trjlen_ri}

\begin{figure*}
          \centering
          \includegraphics[width=\linewidth]{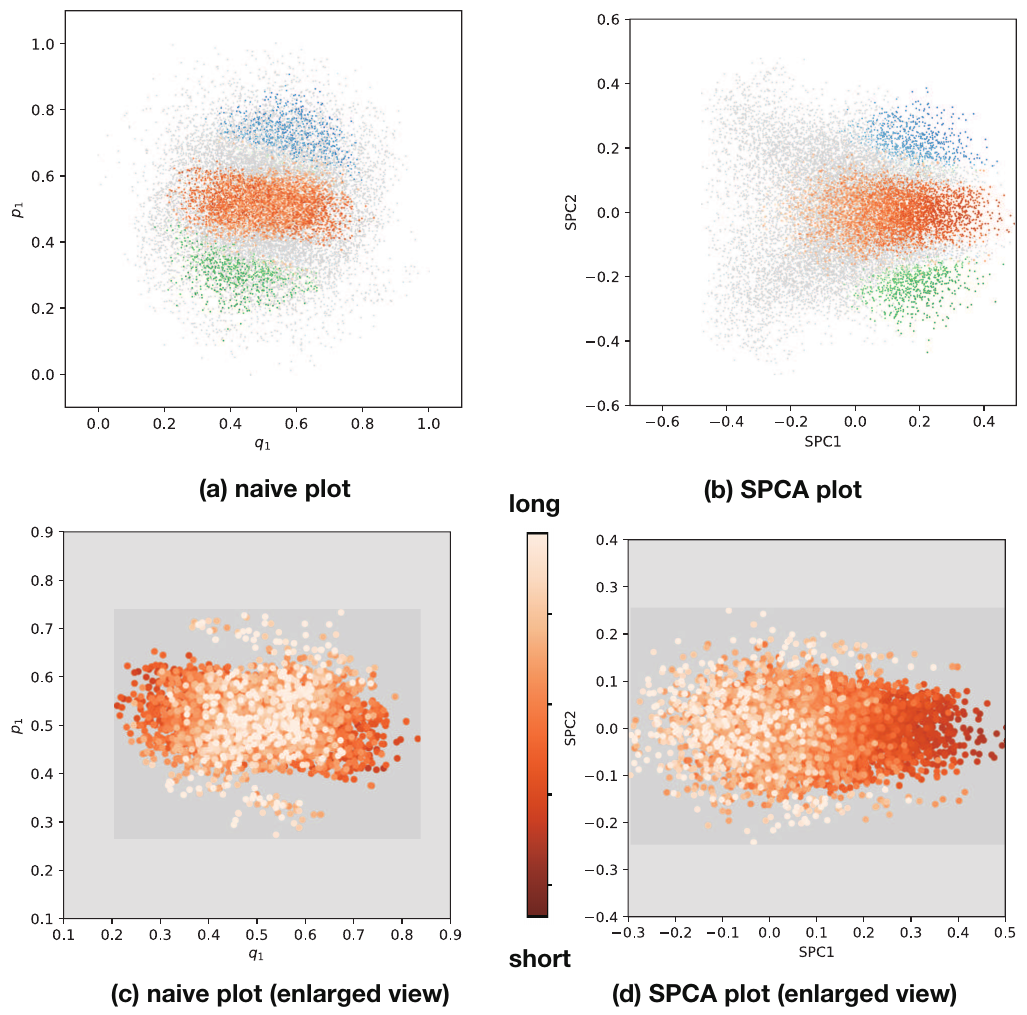}
          \caption{
            Projected RIs on two-dimensional planes with color brightness indicating trajectory passage time.
            The data points are the same as those shown in Fig.~\ref{fig:result_of_projection}.
            The color of each data point indicates the reactivity label of the point---class 0: gray,
            class 1: green, class 2: blue, and class 3: red. The color brightness indicates
            the trajectory passage time; brighter colors represent longer passage times.
            Note that the scale of the brightness gradation is determined for each class.
            Furthermore, to clarify the location of the projected RI boundaries corresponding to
            bright color points, the points are plotted so that the brighter points are more
            in the foreground. Panels (c) and (d) are magnified views of red points (class 3)
            in the central part of panels (a) and (b), respectively.
          }
          \label{fig:len_8dof}
        \end{figure*}

        The passage time of a trajectory through a saddle region is an indicator of a RI boundary
        \cite{Nagahata2020}. Figures~\ref{fig:len_8dof}~(a) and (b), respectively, depict
        the data points in Figs.~\ref{fig:result_of_projection}~(a) and (b) with color brightness
        indicating the passage time. In Fig.~\ref{fig:len_8dof}, the passage time is defined as
        the time to hit the classification condition described in Sec.~\ref{subsec:modelHamiltonian} for
        each trajectory starting from $\Sigma^{\mathrm{sp}}$. We also show magnified views of
        the central regions of class 3 in panels (c) and (d) to make the brightness distribution
        and gradient clear.

        In the SPCA case [panels (b) and (d)], bright points with long trajectory passage time
        accumulate around the boundary of each color cluster. Such bright points likely correspond to
        RI boundaries. On the other hand, bright points scatter inside each color cluster in
        the naive projection case [panels (a) and (c)]. These results also support that
        the proposed algorithm is an effective way to capture RI boundaries compared with
        the naive projection.

\section{Concluding remarks}
  \label{sec:conclusion}

  We have formulated the dimensionality reduction problem of RIs and developed an algorithm to
  solve the problem using SPCA. The effectiveness of the proposed algorithm was examined by
  numerical experiments for many-DoF Hénon-Heiles systems. The numerical results indicate that
  our proposed algorithm is effective in terms of the quality of reactivity prediction and
  the clearness of projected RI boundaries, compared with the naive projection method.

  The algorithm proposed in this paper is a prototype of dimensionality reduction algorithms
  for numerical phase space geometry. Although the SPCA employed in the present study is
  a linear dimensionality reduction method, the SPCA can be extended to nonlinear dimensionality
  reduction method with kernel trick \cite{Scholkopf2002}. Moreover, there are other supervised dimensionality
  reduction techniques, such as partial least squares \cite{wold1975}, canonical correlation
  analysis \cite{Hotelling1936}, and UMAP \cite{McInnes2018a}. 
  Applying these methods may enhance the effectiveness of the dimensionality reduction of RIs.
  In addition, focusing on clustering rather than just dimensionality reduction could lead to 
  an interesting development in discussing reactive islands. The research of Wiggins et al.
  ~\cite{Naik2021} would be informative in this regard.

  Systematic methods for determining appropriate dimensionality $d$ of projection should
  also be developed. In the present study, we determined $d$ based on the cumulative
  contribution rate. However, this is not the unique choice for determining $d$. For instance,
  we can adopt residual variance \cite{Tenenbaum2000} as another criterion for this purpose.
  Model selection criteria such as AIC\cite{Akaike1974} and BIC\cite{Schwarz1978}
  and sparse modeling techniques \cite{Zou2006, Adachi2016} may also be applicable to
  the determination of $d$. Appropriate model selection methods for numerical phase space
  geometry remain open for further investigation.

  The final goal of numerical phase space geometry is to elucidate dynamical reaction mechanisms
  of real chemical systems. Although the present study employs model Hamiltonian systems for
  the proof of concept, our algorithm does not assume analytical expressions of Hamiltonian.
  Therefore, the proposed algorithm can be combined with \textit{ab initio} molecular dynamics
  calculation to compute projected RIs in real molecular systems. Additionally, one can integrate
  the dimensionality reduction algorithm with the efficient trajectory sampling algorithms
  for computing RIs \cite{Nagahata2020, Mizuno2021} (see also Sec.~\ref{sec:boundary-detecton}).
  Although the curse of dimensionality in computing phase space structures can be mitigated
  by dimensionality reduction techniques, the required number of trajectory calculation
  may still be too large to apply the algorithm to real molecular systems of many-DoF.
  This issue may be resolved by using a neural network potential \cite{Manzhos2021} as a surrogate model
  for molecular dynamics simulation. We expect that the further development of numerical methods
  for phase space geometry and their application to real chemical systems will practicalize
  and facilitate the dissemination of chemical reaction analysis from the point of view of 
  dynamical systems theory.

\begin{acknowledgments}
  This work was supported by JSPS KAKENHI, Grant Number 20K15224, Japan (to YM),
  and by JST, the establishment of university fellowships towards the creation 
  of science technology innovation, Grant Number JPMJFS2101 (to RT).
  M.T. is supported by the Research Program of “Dynamic Alliance for 
  Open Innovation Bridging Human, Environment and Materials” in “Network 
  Joint Research Center for Materials and Devices” and a Grant-in-Aid for 
  Scientific Research (C) (No.22654047, No. 25610105, No. 19K03653, and 
  No. 23K03265 ) from JSPS.
\end{acknowledgments}

\bibliography{reference}

\end{document}

\preprint{AIP/123-QED}

\title{Supplementary Material for ``Low-Dimensional Projection of Reactive Islands in Chemical Reaction Dynamics
       Using a Supervised Dimensionality Reduction Method''}

\author{Ryoichi Tanaka}
    \affiliation{Graduate School of Chemical Sciences and Engineering, Hokkaido University, Sapporo 060-8628, Japan}
\author{Yuta Mizuno}
    \affiliation{Graduate School of Chemical Sciences and Engineering, Hokkaido University, Sapporo 060-8628, Japan}
    \affiliation{Research Institute for Electronic Science, Hokkaido University, Sapporo 001-0020, Japan}
    \affiliation{Institute for Chemical Reaction Design and Discovery (WPI-ICReDD), Hokkaido University, Sapporo 001-0021, Japan}
    \email{mizuno@es.hokudai.ac.jp}
\author{Takuro Tsutsumi}
    \affiliation{Graduate School of Chemical Sciences and Engineering, Hokkaido University, Sapporo 060-8628, Japan}
    \affiliation{Department of Chemistry, Faculty of Science, Hokkaido University, Sapporo 060-0810, Japan}
\author{Mikito Toda}
    \affiliation{Graduate School of Information Science, University of Hyogo, Kobe, 650-0047, Japan}
    \affiliation{Research Institute for Electronic Science, Hokkaido University, Sapporo 001-0020, Japan}
    \affiliation{Faculty Division of Natural Sciences, Research Group of Physics, Nara Women’s University,  Nara 630‐8506, Japan}
\author{Tetsuya Taketsugu}
    \affiliation{Graduate School of Chemical Sciences and Engineering, Hokkaido University, Sapporo 060-8628, Japan}
    \affiliation{Institute for Chemical Reaction Design and Discovery (WPI-ICReDD), Hokkaido University, Sapporo 001-0021, Japan}
    \affiliation{Department of Chemistry, Faculty of Science, Hokkaido University, Sapporo 060-0810, Japan}
\author{Tamiki Komatsuzaki}
    \affiliation{Graduate School of Chemical Sciences and Engineering, Hokkaido University, Sapporo 060-8628, Japan}
    \affiliation{Research Institute for Electronic Science, Hokkaido University, Sapporo 001-0020, Japan}
    \affiliation{Institute for Chemical Reaction Design and Discovery (WPI-ICReDD), Hokkaido University, Sapporo 001-0021, Japan}
    \affiliation{SANKEN, Osaka University, Ibaraki 567-0047, Japan}

\date{\today}

\maketitle

\tableofcontents

\section{Projection Results: the 4, 6, and 10-DoF Cases}

\begin{figure*}[h!]
    \centering
    \includegraphics[width=\linewidth]{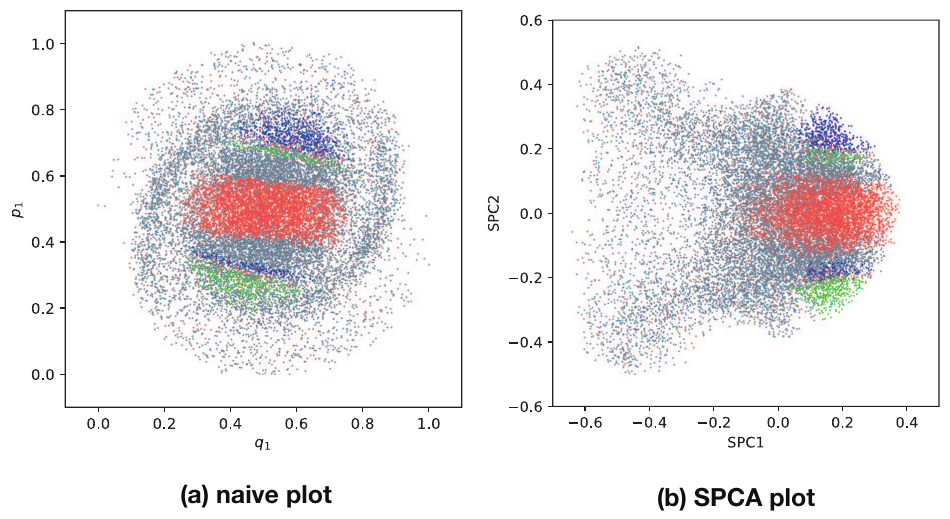}
    \caption{
      Projected RIs on two-dimensional planes for the 4-DoF Hénon-Heiles system.  
      For more details, see Fig.~5 in the main paper.
    }
    \label{fig:result_of_projection_4dof}
    \end{figure*}

\begin{figure}[h!]
    \centering
    \includegraphics[width=10cm]{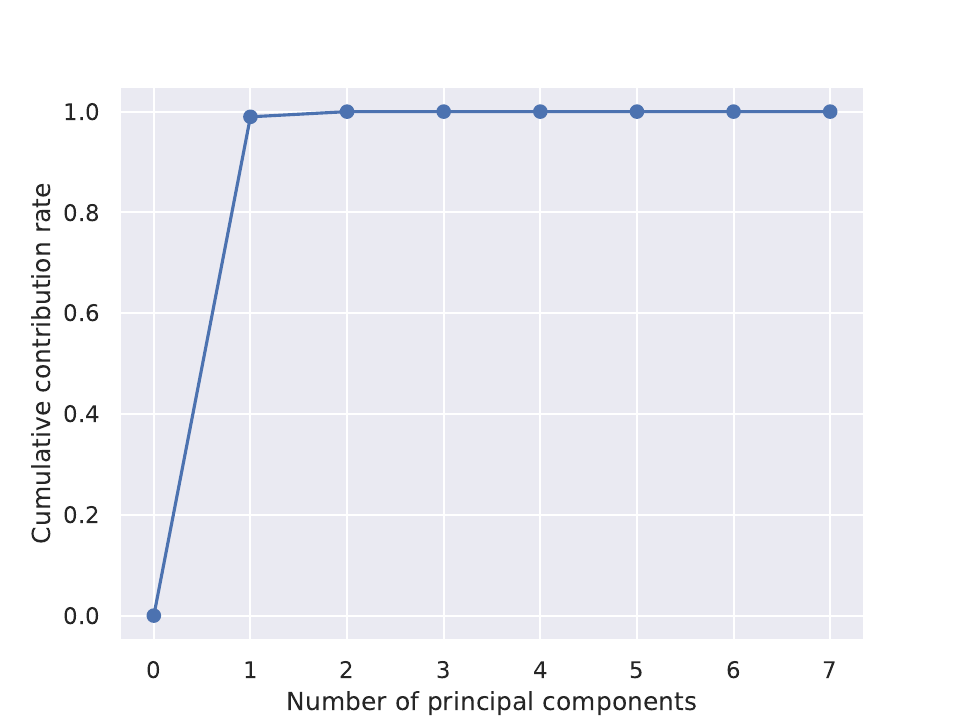}
    \caption{
      Cumulative contribution rates of the SPCA result for the 4-DoF system.
      The top three eigenvalues are $9.8 \times 10^{-1}$,
      $1.0 \times 10^{-2}$, $1.1 \times 10^{-4}$, while the remaining eigenvalues
      are less than $10^{-17}$.
      For more details, see Fig.~6 in the main paper.
    }
    \label{fig:cum_cont_rate_4dof}
    \end{figure}

\begin{figure*}[h!]
    \centering
    \includegraphics[width=\linewidth]{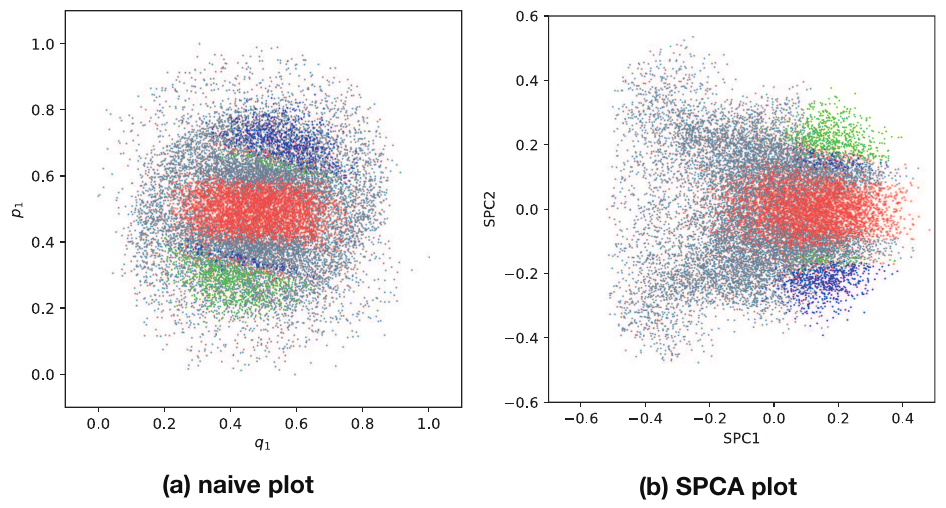}
    \caption{
      Projected RIs on two-dimensional planes for the 6-DoF Hénon-Heiles system.   
      For more details, see Fig.~5 in the main paper.
    }
    \label{fig:result_of_projection_6dof}
    \end{figure*}

\begin{figure}[h!]
    \centering
    \includegraphics[width=10cm]{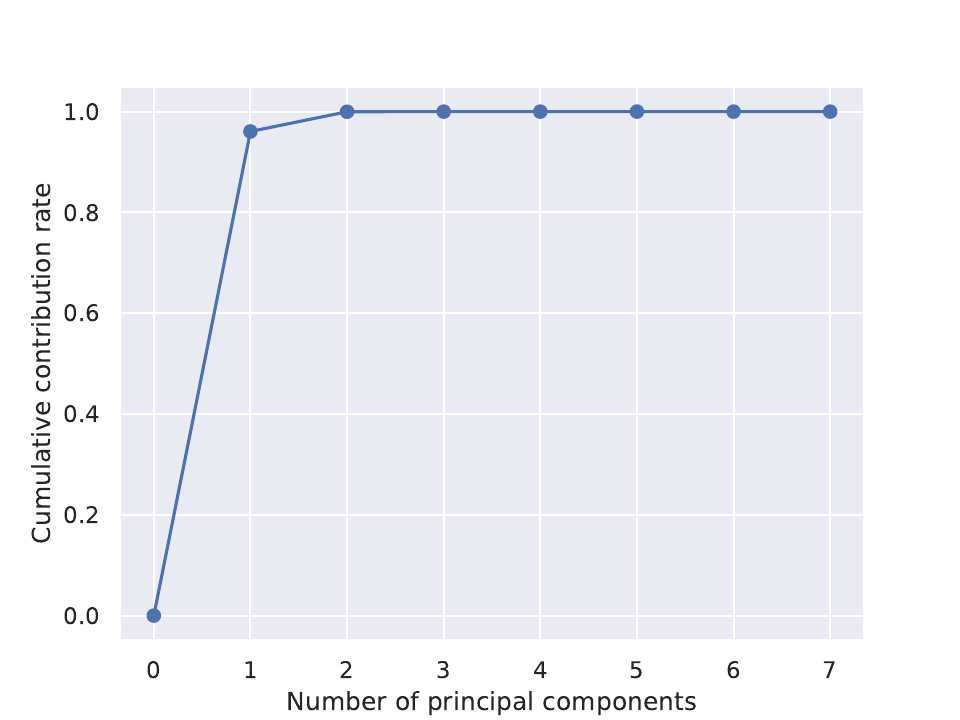}
    \caption{
      Cumulative contribution rates of the SPCA result for the 6-DoF system.
      The top three eigenvalues are $9.5 \times 10^{-1}$,
      $3.9 \times 10^{-2}$, $3.7 \times 10^{-4}$, while the remaining eigenvalues
      are less than $10^{-16}$.
      For more details, see Fig.~6 in the main paper.
    }
    \label{fig:cum_cont_rate_6dof}
    \end{figure}

\begin{figure*}[h!]
    \centering
    \includegraphics[width=\linewidth]{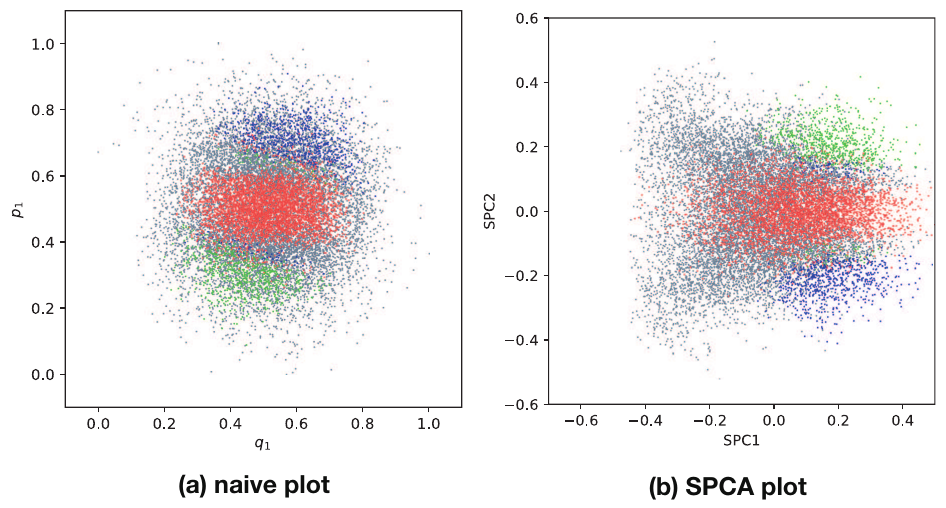}
    \caption{
      Projected RIs on two-dimensional planes for the 10-DoF Hénon-Heiles system.    
      For more details, see Fig.~5 in the main paper.
    }
    \label{fig:result_of_projection_10dof}
    \end{figure*}

\begin{figure}[H]
    \centering
    \includegraphics[width=10cm]{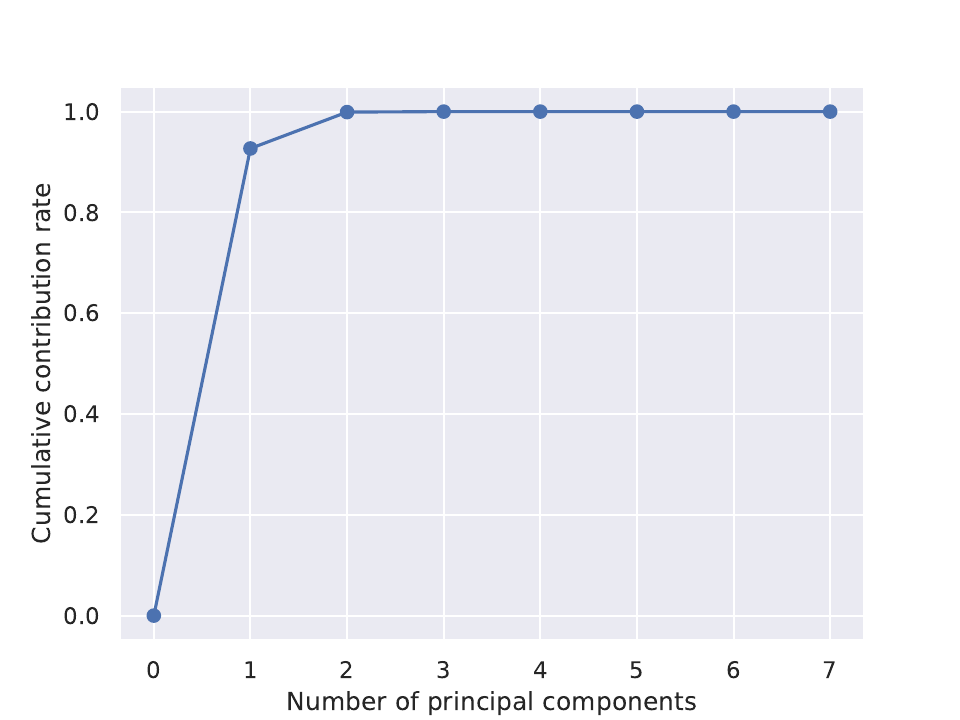}
    \caption{
      Cumulative contribution rates of the SPCA result for the 10-DoF system.
      The top three eigenvalues are $9.2 \times 10^{-1}$,
      $7.0 \times 10^{-2}$, $8.6 \times 10^{-4}$, while the remaining eigenvalues
      are less than $10^{-16}$.
      For more details, see Fig.~6 in the main paper.
    }
    \label{fig:cum_cont_rate_10dof}
    \end{figure}

    This supplementary section presents (1) reactive islands (RIs) projected onto
    two-dimensional planes by the naive and SPCA projections
    (Figs.~\ref{fig:result_of_projection_4dof}, \ref{fig:result_of_projection_6dof},
    and \ref{fig:result_of_projection_10dof}), (2) the cumulative contribution rates
    of SPCA (Figs.~\ref{fig:cum_cont_rate_4dof}, \ref{fig:cum_cont_rate_6dof}, and
    \ref{fig:cum_cont_rate_10dof}), and (3) the compositions of the top-three
    principal components (Tables~\ref{tab:construction_pc_4dof}, 
    \ref{tab:construction_pc_6dof}, and \ref{tab:construction_pc_10dof}),
    for the 4, 6, and 10-DoF systems.
    Note that all numerical values in this paper are truncated after the third significant digit.

    These results for the 4, 6, and 10-DoF systems exhibit similar to
    those for the 8-DoF system shown in Sec.~IV~B
    of the main text:  (1) For all systems, compared to the RIs projected
    by the naive projection, those projected by our SPCA algorithm appear to be
    less mixed with other regions. (2) The cumulative contribution rates indicate
    that the projection onto the top-two principal components preserves the HSIC
    value sufficiently. (3) The $p_2$ is the dominant component of the first principal
    component, while $p_1$ and $q_1$ is those of the second principal component.

\clearpage


\clearpage


\clearpage


    \begin{table}[h!]
      \caption{
        Compositions of the top-three principal components (PCs) for the 4-DoF system.
        Since the data points are on the hypersurface $q_2=0$, all the contributions
        of $q_2$ are zero. 
      }
      \label{tab:construction_pc_4dof}
      \centering
      \begin{tabular}{lrrrrrrrr}
        \hline
        & $q_1$ & $p_1$ & $q_2$ & $p_2$ & $q_3$ & $p_3$ & $q_4$ & $p_4$  \\
        \hline
        PC1 &  -0.00095 & -0.012 & 0.0 & 0.99 & 0.0098 & 0.012 & 0.039 \\   
        PC2 & -0.34 & -0.93 & 0.0 & -0.012 & 0.029 & -0.0044 & 0.010 \\
        PC3 & 0.59 & -0.19 & 0.0 & -0.014 & 0.64 & -0.25 & 0.32  \\
        \hline 
      \end{tabular}
    \end{table}

    \begin{table}[h!]
      \caption{
        Compositions of the top-three principal components (PCs) for the 6-DoF system.
        Since the data points are on the hypersurface $q_2=0$, all the contributions
        of $q_2$ are zero.
      }
      \label{tab:construction_pc_6dof}
      \centering
      \begin{tabular}{lrrrrrrrrrrrr}
        \hline
        & $q_1$ & $p_1$ & $q_2$ & $p_2$ & $q_3$ & $p_3$ & $q_4$ & $p_4$ & $q_5$ & $p_5$ & $q_6$ & $p_6$ \\
        \hline
        PC1 & -0.015 & -0.034 & 0.0 & -0.99 & -0.016 & -0.016 & -0.010 & 0.017 & -0.0021 &  0.010 & 0.016 & -0.0042 \\
        PC2 & -0.34 & -0.93 & 0.0 & 0.037 & -0.0039 & 0.010 & 0.026 & -0.017 & -0.0047 & -0.016 & 0.0042 &  0.012 \\
        PC3 &  0.22 & -0.075 & 0.0 & -0.013 & 0.21 & 0.55 & -0.15 & -0.40 & -0.54 & -0.070 & 0.25 & -0.20 \\
        \hline 
      \end{tabular}
    \end{table}

    \begin{table}[h!]
      \caption{
        Compositions of the top-three principal components (PCs) for the 10-DoF system.
        Since the data points are on the hypersurface $q_2=0$, all the contributions
        of $q_2$ are zero.
      }
      \label{tab:construction_pc_10dof}
      \centering
      \begin{tabular}{lrrrrrrrrrr}
        \hline
        & $q_1$ & $p_1$ & $q_2$ & $p_2$ & $q_3$ & $p_3$ & $q_4$ & $p_4$ & $q_5$ & $p_5$ \\
        \hline
        PC1 & -0.0057 & 0.0040 & 0.0 & -0.99 & 0.010 & -0.011 & -0.035 & -0.022 & -0.020 & -0.031 \\
        PC2 & -0.35 & -0.92 & 0.0 & -0.0030 & -0.014 &  0.0085 & 0.0041 &  0.012 &  0.027 & 0.0012 \\
        PC3 &  0.20 & -0.051 & 0.0 & -0.058 & -0.37 & -0.060 & 0.37 &  0.39 &  0.24 & 0.36 \\
        \hline \hline 
        & $q_6$ & $p_6$ & $q_7$ & $p_7$ & $q_8$ & $p_8$ & $q_9$ & $p_9$ & $q_{10}$ & $p_{10}$ \\
        \hline
        PC1 &  0.028 & 0.0060 &  0.0091 &   -0.017 & -0.011 &  0.0065 &  0.010 &  0.022 & -0.0093 & -0.0082 \\
        PC2 & -0.0060 & 0.014 & 0.011 & -0.010 & 0.014 & -0.025 & -0.020 &  0.020 & -0.025 &  0.044 \\
        PC3 & -0.12 & -0.27 & 0.19 & 0.13 & 0.098 & 0.048 & -0.32 & -0.17 & 0.046 & 0.18 \\
        \hline
      \end{tabular}
    \end{table}

\clearpage

\section{Reactivity Prediction by the K-Nearest Neighbor Method}
    In Sec.~IV~C in the main part of the paper, we evaluate
    the effectiveness of the proposed algorithm in terms of the quality of
    the reactivity label prediction. The label predictions are conducted
    with the k-nearest neighbor method. To avoid complications, original results
    of reactivity predictions are not shown in the main part of the paper.
    The result of predictions for all simulation (4, 6, 8, 10-DoF systems)
    are shown in this supplementary section. 
    
    \begin{figure*}[h!]
    \centering
    \includegraphics[width=\linewidth]{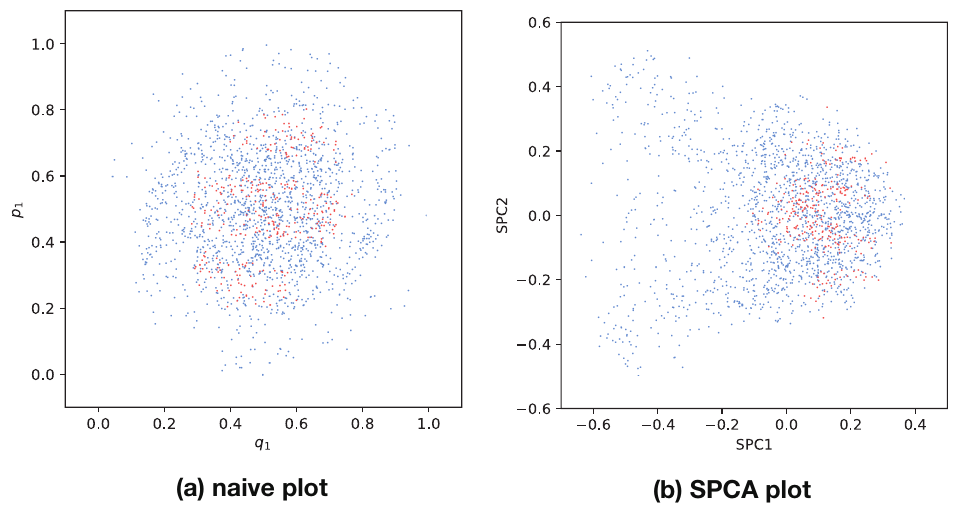}
    \caption{
      The correctness of the k-nearest neighbor prediction for 4-DoF system.
      The parameter $k$ is $k=5$, and the test data and training data were randomly split 
      in a ratio of 1:9.
      For detailed information on the number of data points, see Tab.~\ref{tab:number_of_datapoints}.
      Blue and red points indicate correct and incorrect predictions, respectively.
    }
    \end{figure*}

    \begin{figure*}[h!]
    \centering
    \includegraphics[width=\linewidth]{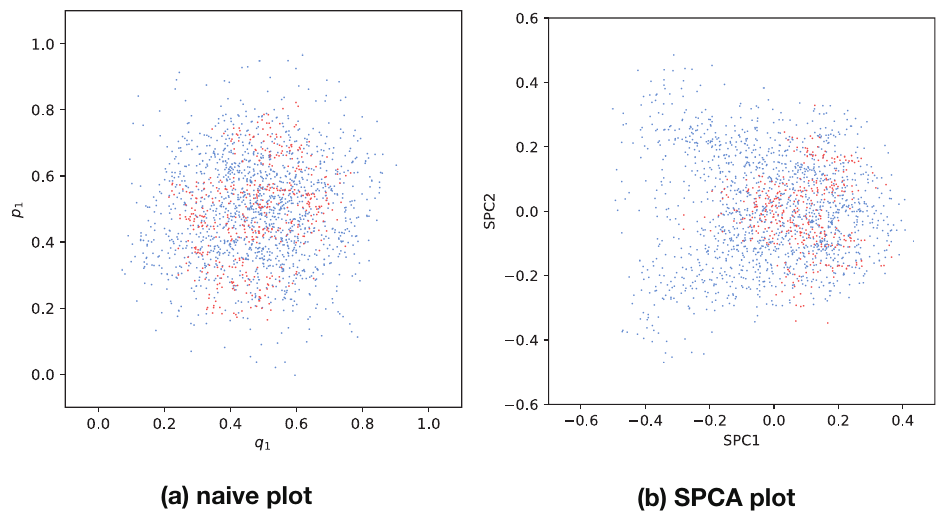}
    \caption{
      The correctness of the k-nearest neighbor prediction for 6-DoF system.
      The parameter $k$ is $k=5$, and the test data and training data were randomly split 
      in a ratio of 1:9.
      For detailed information on the number of data points, see Tab.~\ref{tab:number_of_datapoints}.
      Blue and red points indicate correct and incorrect predictions, respectively.
    }
    \end{figure*}

    \begin{figure*}[h!]
    \centering
    \includegraphics[width=\linewidth]{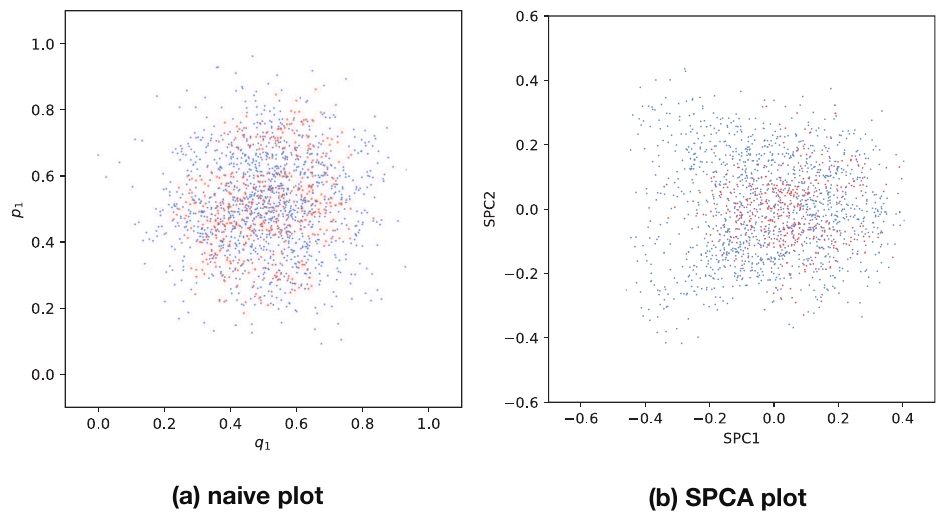}
    \caption{
      The correctness of the k-nearest neighbor prediction for 8-DoF system.
      The parameter $k$ is $k=5$, and the test data and training data were randomly split 
      in a ratio of 1:9.
      For detailed information on the number of data points, see Tab.~\ref{tab:number_of_datapoints}.
      Blue and red points indicate correct and incorrect predictions, respectively.
    }
    \end{figure*}

    \begin{figure*}[h!]
    \centering
    \includegraphics[width=\linewidth]{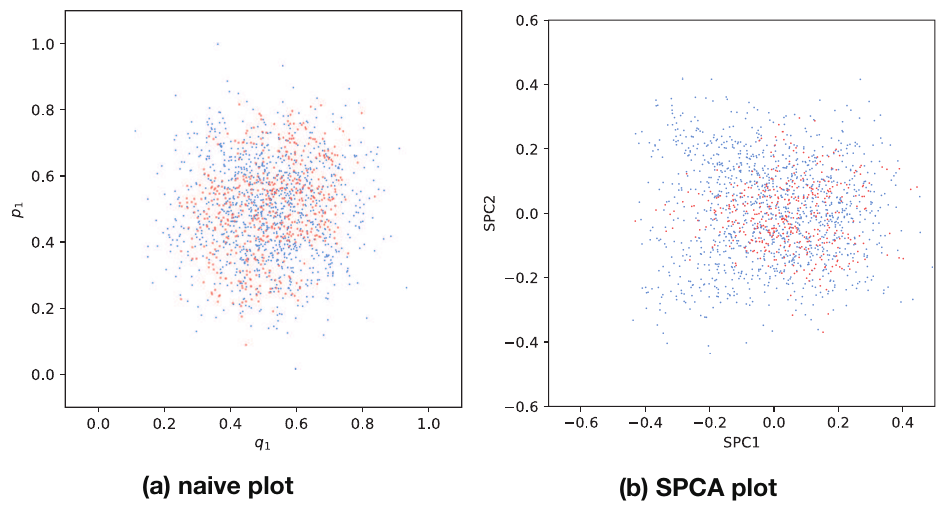}
    \caption{
      The correctness of the k-nearest neighbor prediction for 10-DoF system.
      The parameter $k$ is $k=5$, and the test data and training data were randomly split 
      in a ratio of 1:9.
      For detailed information on the number of data points, see Tab.~\ref{tab:number_of_datapoints}.
      Blue and red points indicate correct and incorrect predictions, respectively.
    }
    \end{figure*}

\clearpage

\section{Why the Macro-F1 Score Is Adopted}

    In Sec.~\ref{sec:label-prediction} in the main part of the paper,
    we evaluate the quality of the reactivity label prediction by
    the k-nearest neighbor method with the macro-F1 score.
    This supplementary section describes the reason why the criterion is adopted.

    The reactivity prediction is a rare-event prediction. As shown in
    Tab.~\ref{tab:number_of_datapoints}, reaction events (class 1--3) are rare
    compared with non-reactive events (class 0).
    
    \begin{table}[h!]
        \caption{The number of data points in test data for each label for each DoF.
        }
        \label{tab:number_of_datapoints}
        \centering
        \begin{tabular}{lrrrr}
          \hline
                    & 4-DoF & 6-DoF & 8-DoF & 10-DoF  \\
            \hline \hline
            class 0 & 13909 & 12656 & 11034 & 10243 \\
            \hline
            class 1 & 534 & 773 & 725 & 765 \\
            \hline
            class 2 & 545 & 726 & 803 & 761 \\
            \hline
            class 3 & 4377 & 4245 & 4511 & 4438 \\
            \hline \hline
            total   & 19365 & 18400 &  17073 & 16207 \\
          \hline  
        \end{tabular}
    \end{table}

    The prediction accuracy, a common indicator of prediction performance,
    is not necessarily suitable for evaluating such a rare-event prediction task.
    Consider a binary classification task between A and B and suppose A is rare.
    A classifier that assigns all objects to B may have a high prediction accuracy;
    it always missclassifies A as B but such missclassification rarely occurs
    due to the rareness of A. In other words, most of the predictions are
    true negative. The prediction accuracy is defined as
    \begin{equation}
        \mathrm{Accuracy} = \frac{\mathrm{TP}+\mathrm{TN}}
                                 {\mathrm{TP}+\mathrm{FP}+\mathrm{TN}+\mathrm{FN}},
    \end{equation}
    where TP, FP, TN, and FN indicate the number of true positive, false positive,
    true negative, and false negative, respectively. When TN is sufficiently larger
    than the others, the accuracy is close to one. Thus, an appropriate performance
    indicator of rare-event predictions should be independent of TN.

    The F1 score is defined as the harmonic mean of the recall and precision,
    which are independent of TN:
    \begin{align*}
        F_1 = \frac{2 \cdot \mathrm{Recall} \cdot \mathrm{Precision}}
                   {\mathrm{Recall} + \mathrm{Precision}},
    \end{align*}
    where
    \begin{align*}
        \mathrm{Recall} &= \frac{\mathrm{TP}}{\mathrm{TP}+\mathrm{FN}}, \\
        \mathrm{Precision} &= \frac{\mathrm{TP}}{\mathrm{TP}+\mathrm{FP}}.
    \end{align*}
    According to the above discussion, the F1 score is more suitable than
    the prediction accuracy as the performance indicator for binary classification
    tasks with rare-events.

    The F1 score for a multiclass classification is defined by an average of
    the F1 scores for elementary binary classification class. For example,
    in our case, the classification between class $l$ and the others is
    an elementary binary classification. There are two standard averaging method,
    macro-average and micro-average \cite{Manning2008, Sokolova2009, Takahashi2022}. 
    The macro-average is the arithmetic mean of the F1 scores for binary classifications:
    \begin{equation}
        \mathrm{macro\mathchar`-}F_1 = \frac{\sum_{l=0}^3 F_1(\mathrm{class} \; l)}{4},
    \end{equation}
    where $F_1(\mathrm{class} \; l)$ is the F1 score of the binary classification
    between class $l$ and the others. On the other hand,
    in the calculation of micro-average, firstly micro-recall and micro-precision are
    calculated:
    \begin{align*}
        \mathrm{micro\mathchar`-Recall} &= \frac{\sum_{l=0}^3 \mathrm{TP}_l}{\sum_{l=0}^3 (\mathrm{TP}_l+\mathrm{FN}_l)}, \\
        \mathrm{micro\mathchar`-Precision} &= \frac{\sum_{l=0}^3 \mathrm{TP}_l}{\sum_{l=0}^3 (\mathrm{TP}_l+\mathrm{FP}_l)}.
    \end{align*}
    Here, $\mathrm{TP}_l$, $\mathrm{FP}_l$, $\mathrm{TN}_l$, and $\mathrm{FN}_l$ indicate the number 
    of true positive, false positive, true negative, and false negative, for each class $l$ respectively.
    Then micro-$F_1$ score is derived from the two values:
    \begin{align*}
        \mathrm{micro\mathchar`-}F_1 = \frac{2 \cdot \mathrm{micro\mathchar`-Recall} \cdot \mathrm{micro\mathchar`-Precision}}
                   {\mathrm{micro\mathchar`-Recall} + \mathrm{micro\mathchar`-Precision}}.
    \end{align*}
    The micro-average is influenced by the imbalance in data for each class.
    To focus on the performance of predicting rare reaction events, we adopt
    the macro-F1 score in the present paper.

\clearpage

\section{F1 Score Ratio Along $k$}
\begin{figure}[h!]
  \centering
  \includegraphics[width=\linewidth]{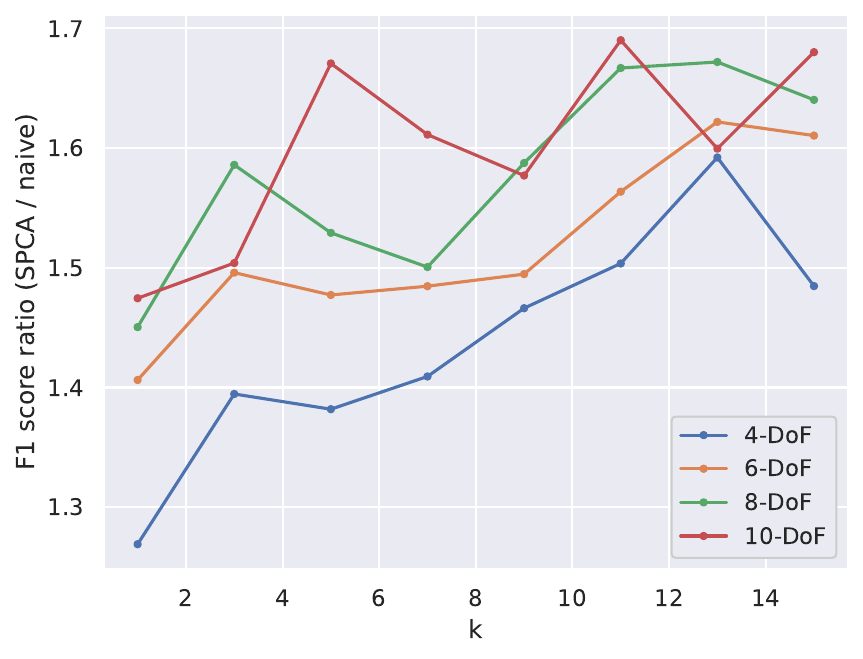}
  \caption{
    The ratio of the F1 value (SPCA/naive) for each $k$ value, computed for each DoF system.
  }
  \label{fig:f1_k_ratio}
\end{figure}

    In Sec.~IV~C in the main part of the paper,
    the macro-F1 score ratio was calculated for the case of
    $k =$ 3, 5, 7, 9, 11, 13, and 15 for each-DoF system,
    but the detailed data are not shown to avoid complications.
    Figure~\ref{fig:f1_k_ratio} shows the ratio of the F1 score for each $k$.
    For all $k$ and DoF, the F1 score ratio is larger than one,
    which shows the effectiveness of the SPCA method.

\clearpage

\section{Label Separation: the 4, 6, and 10-DoF Cases}
\begin{figure*}[h!]
    \centering
    \includegraphics[width=\linewidth]{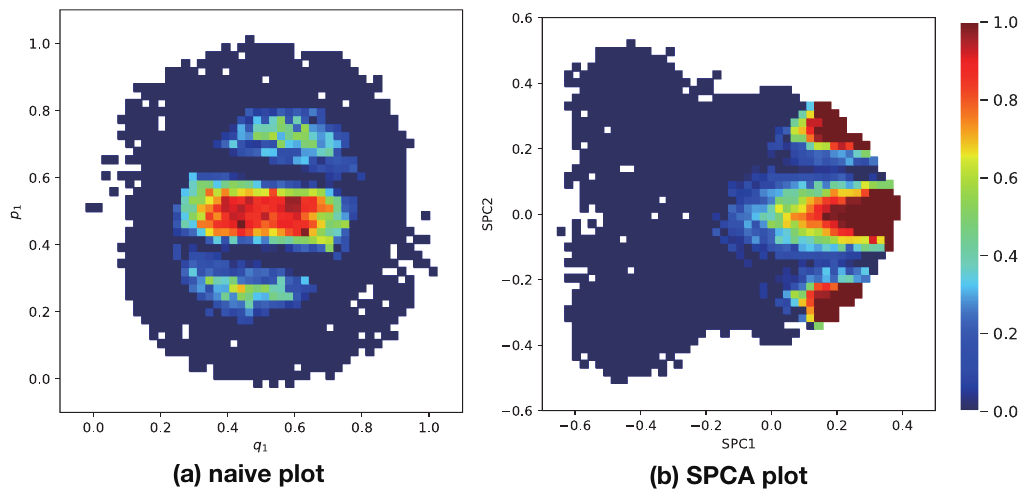}
    \caption{
      Heatmaps of the certainty of RI attribution for 4-DoF system.
      This heatmaps are calculated from 19365 data points.
      For more details, see Fig.~\ref{fig:upper_dec_8dof} in the main paper.
    }
    \label{fig:upper_dec_4dof}
    \end{figure*}

\begin{figure*}[h!]
    \centering
    \includegraphics[width=\linewidth]{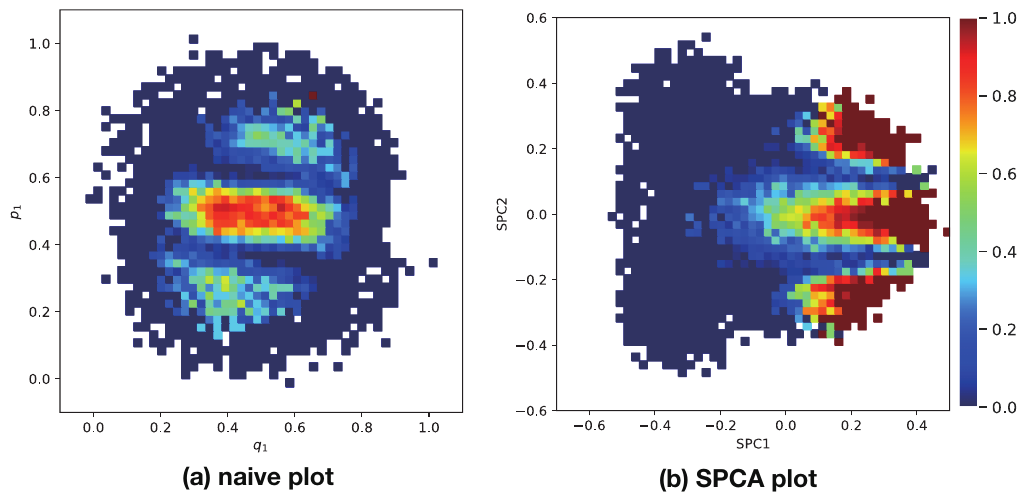}
    \caption{
      Heatmaps of the certainty of RI attribution for 6-DoF system.
      This heatmaps are calculated from 18400 data points.
      For more details, see Fig.~\ref{fig:upper_dec_8dof} in the main paper.
    }
    \label{fig:upper_dec_6dof}
    \end{figure*}

\begin{figure*}[h!]
    \centering
    \includegraphics[width=\linewidth]{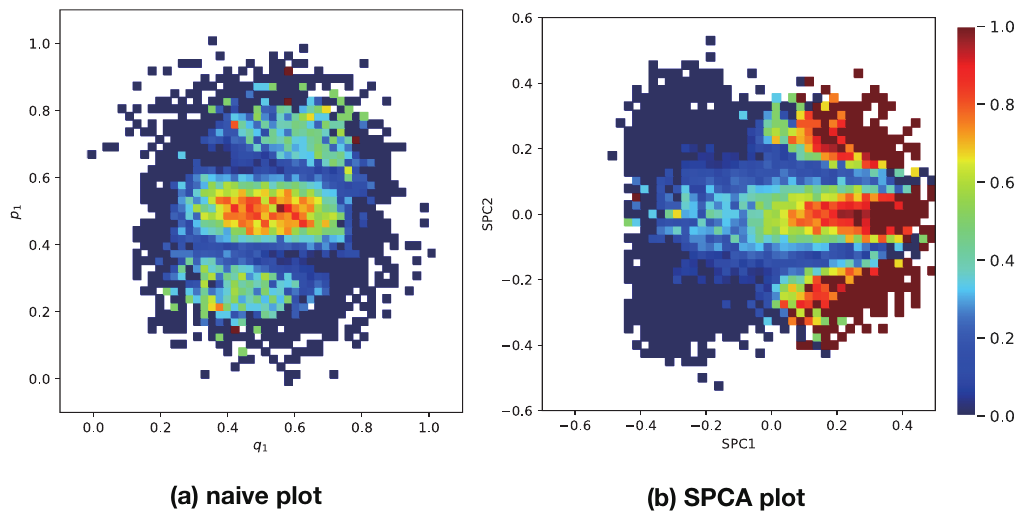}
    \caption{
      Heatmaps of the certainty of RI attribution for 10-DoF system.
      This heatmaps are calculated from 16207 data points.
      For more details, see Fig.~\ref{fig:upper_dec_8dof} in the main paper.
    }
    \label{fig:upper_dec_10dof}
    \end{figure*}

    In Sec.~IV~D in the main part of the paper,
    we plotted the heatmaps to visualize the fuzziness of projected RIs
    for the 8-DoF system. Here, we show the result of the 4, 6, and 10-DoF systems
    in Figs.~\ref{fig:upper_dec_4dof}--\ref{fig:upper_dec_10dof}.

    Here, we briefly explain the grid generation process. Firstly, the range is set
    such that the maximum and minimum values are the centers of the end grids. 
    Secondly, for grids containing one or more points, the grid width is determined
    such that the average number of points within the grid is 15 points.
    
\clearpage

\section{Trajectory Passage Time: the 4, 6, and 10-DoF Cases}
\begin{figure*}[h!]
    \centering
    \includegraphics[width=\linewidth]{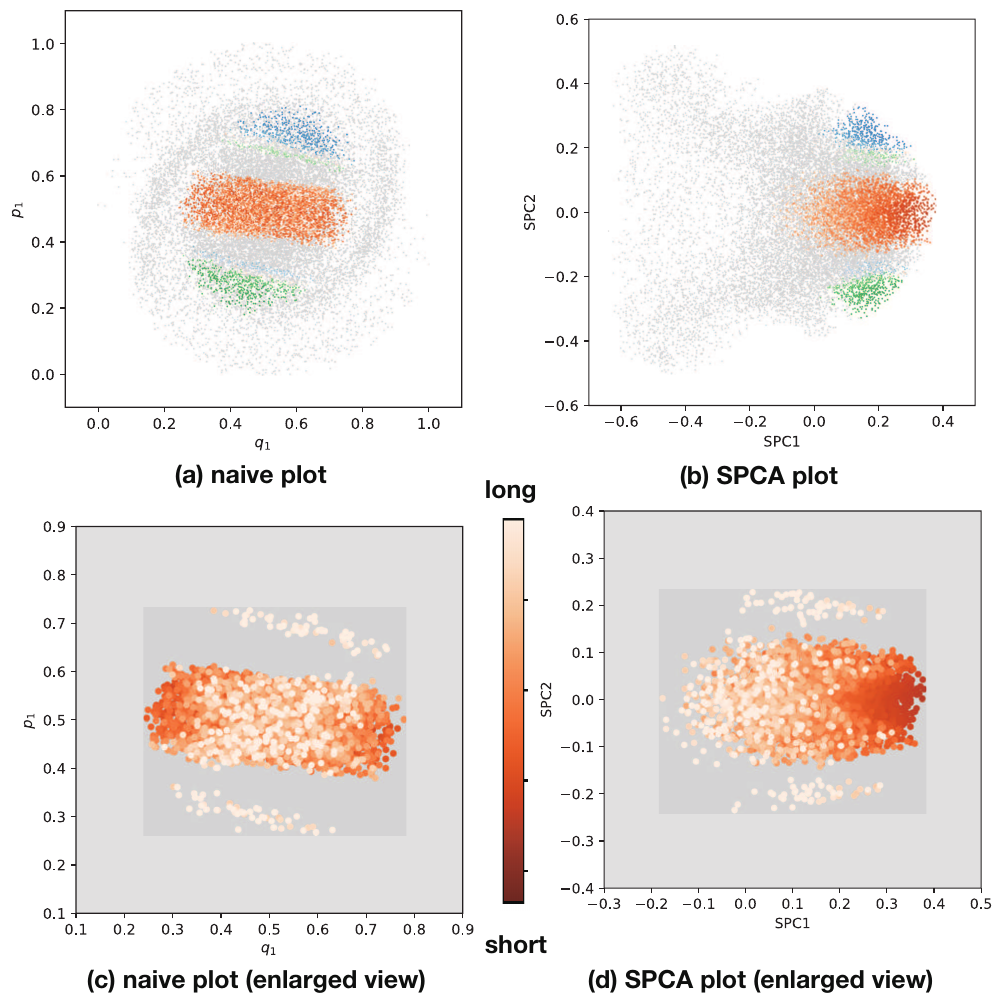}
    \caption{
        Projected RIs on two-dimensional planes with color brightness indicating trajectory passage time for 4-DoF system.
        For more details, see Fig.~\ref{fig:len_8dof} in the main paper.
    }
    \label{fig:len_4dof}
    \end{figure*}

\begin{figure*}[h!]
    \centering
    \includegraphics[width=\linewidth]{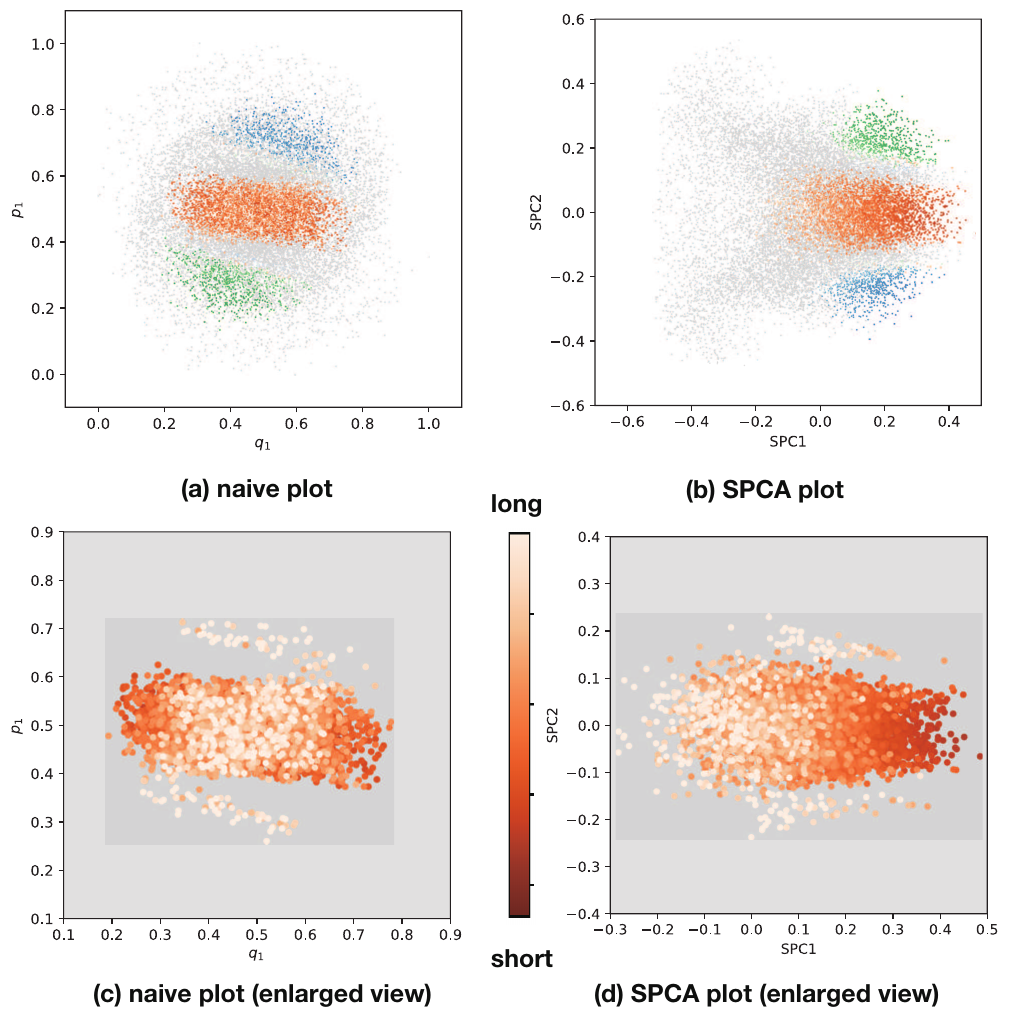}
    \caption{
        Projected RIs on two-dimensional planes with color brightness indicating trajectory passage time for 6-DoF system.
        For more details, see Fig.~\ref{fig:len_8dof} in the main paper.
    }
    \label{fig:len_6dof}
    \end{figure*}

\begin{figure*}[h!]
    \centering
    \includegraphics[width=\linewidth]{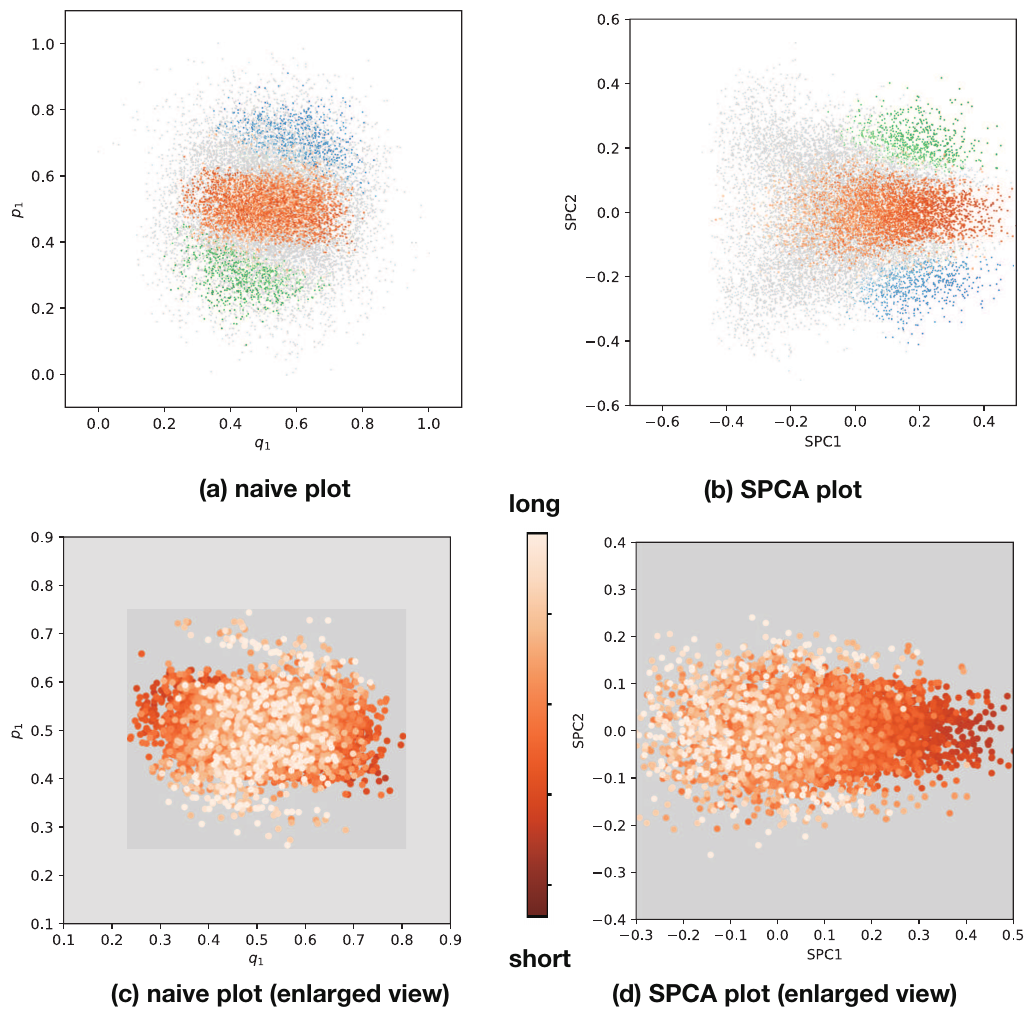}
    \caption{
        Projected RIs on two-dimensional planes with color brightness indicating trajectory passage time for 10-DoF system.
        For more details, see Fig.~\ref{fig:len_8dof} in the main paper.
    }
    \label{fig:len_10dof}
    \end{figure*}

    In Sec.~\ref{sec:subsub_trjlen_ri} in the main part of the paper,
    to show the reactivity boundary in the projected space,
    data points in low dimensional space were colored by the passage time of the trajectories
    for 8-DoF system in Fig.~\ref{fig:len_8dof} in the main part of the paper.
    In this section, we also show the result of 4, 6, 10-DoF systems in
    Figs.~\ref{fig:len_4dof}--\ref{fig:len_10dof}.
    
\clearpage

\clearpage

\bibliography{reference}